\documentclass{article}
\usepackage{epsfig}
\usepackage{amssymb}
\usepackage{graphicx}
\usepackage{epsf}
\usepackage{calrsfs}
\usepackage{txfonts}

\begin{document}

%%% math shortcuts
\newcommand{\be}{\begin{equation}}
\newcommand{\ee}{\end{equation}}
\newcommand{\bea}{\begin{eqnarray}}
\newcommand{\eea}{\end{eqnarray}}
\newcommand{\rr}{{\bf r}}
\newcommand{\rt}{\rr^\perp}
\newcommand{\kk}{{\bf k}}
\newcommand{\kkt}{\kk^\perp}
\newcommand{\p}{{\bf p}}
\newcommand{\q}{{\bf q}}
\newcommand{\qt}{\q^\perp}
\newcommand{\X}{({\bf x})}
\newcommand{\Y}{({\bf y})}
\newcommand{\x}{{\bf x}}
\newcommand{\ff}{{\bf f}}
\newcommand{\uu}{{\bf u}}
\newcommand{\xx}{{\bf x}}
\newcommand{\EE}{{\bf E}}
\newcommand{\VV}{{\bf V}}
\newcommand{\y}{{\bf y}}
\newcommand{\U}{{\bf u}}
\newcommand{\w}{{\bf \omega}}
\newcommand{\D}{{\bf \nabla}}
\newcommand{\W}{\omega_\kk}
\newcommand{\za}{\alpha}
\newcommand{\zb}{\beta}
\newcommand{\zd}{\delta}
\newcommand{\zg}{\gamma}
\newcommand{\zl}{\lambda}
\newcommand{\zs}{\sigma}
\newcommand{\zt}{\tau}
\newcommand{\zE}{I\hskip-3.7pt E}
\newcommand{\zN}{I\hskip-3.4pt N}
\newcommand{\zR}{I\hskip-3.4pt R}
\newcommand{\zZ}{Z\hskip-7pt Z}
\newcommand{\zw}{\omega}
\newcommand{\zW}{\Omega}
\newcommand{\zC}{{\mathbb C}}
\newcommand{\zD}{{\Delta}}
\newcommand{\zG}{{\Gamma}}
\newcommand{\veps}{\varepsilon}
\newcommand{\OM}{({\bf ***\ldots***})}
\newcommand{\EM}{({\bf $\leftarrow$***})}
\newcommand{\BM}{({\bf ***$\rightarrow$})}
\newcommand{\noi}{\noindent}

\newcommand {\bdm} {\begin{displaymath}}
\newcommand {\edm} {\end{displaymath}}
\newcommand {\ba}  {\begin{array}}
\newcommand {\ea}  {\end{array}}
\newcommand {\mapx} {\Phi_{\epsilon}}
\newcommand {\Hx}    {{\mathcal H}_{\hat x}}
\newcommand {\muje} {\mu_j(\eps)}
\newcommand {\ugns}[1] {{\bf u}^{gns}(#1)}
\newcommand {\ugnsi}[1] {{\bf u}^{gns}_i(#1)}
\newcommand {\uns}[1] {{\bf u}^{ns}(#1)}
\newcommand {\parno} {\par\noindent}
\newcommand {\para} [1]{ \par\noindent {\bf #1} \par\noindent}
\newcommand {\gam} [1]{\gamma_{{\vec k}_{#1}}}
\newcommand {\norgq} [1] {| \gamma_{{\vec k}_{#1}} |^2}

\newcommand{\td}{T_r}
\newcommand{\ts}{T_\ell}
\newcommand{\xt}{X_t}
\newcommand{\ft}{f_t}
\newcommand{\Ft}{\mathcal{F}_t}
\newcommand{\vt}{\mathcal{V}_t}
\newcommand{\kt}{\mathcal{K}_t}
\newcommand{\rot}{\rho_t}
\newcommand{\cL}{{\tt L}}
\newcommand{\cT}{\mathcal{T}^{(N)}}
\newcommand{\kT}{\mathcal{T}}
\newcommand{\cTa}{\mathcal{T}^\ast}
\newcommand{\se}{{(s)}}
\newcommand{\zP}{\mbox{P}}
\newcommand{\zM}{\mbox {M}}
\newcommand{\pp}{p^+}
\newcommand{\cmin}{\underline\mathcal{C}}
\newcommand{\cmed}{\mathcal{C}_{av}}
\newcommand{\cmax}{\bar\mathcal{C}}

%\renewcommand{\OM}{} \renewcommand{\EM}{} \renewcommand{\BM}{}

%\begin{frontmatter}

\begin{center}
{\bf \Large Anomalies, absence of local equilibrium and universality
in 1-d particles systems}\\
\vspace{1cm}
{\large C. Giberti}\\
{\small Dipartimento di Scienze e Metodi dell'Ingegneria,
Universit\`a di Modena e Reggio~E., Via Amendola 2, Pad. Morselli}\\{\small I-42122 Reggio E., Italy}\\
{\small {e-mail: {\em claudio.giberti@unimore.it}}}\\
\vspace{.5cm}
{\large L. Rondoni}\\
{\small Dipartimento di Matematica, Politecnico di Torino, and INFN,
Corso Duca degli Abruzzi 24, I-10129 Torino, Italy}\\
{\small {e-mail: {\em lamberto.rondoni@polito.it}}}\\
\end{center}
\begin{abstract}
One dimensional systems are under intense investigation, both from theoretical and
experimental points of view, since they have rather peculiar characteristics which
are of both conceptual and technological interest. We analyze the dependence of the
behaviour of one dimensional, time reversal invariant, nonequilibrium systems on
the parameters defining their microscopic dynamics.
In particular, we consider chains of identical oscillators interacting via hard core
elastic collisions and harmonic potentials, driven by boundary Nos\'e-Hoover
thermostats. Their behaviour mirrors qualitatively that of stochastically
driven systems, showing that anomalous properties are typical of physics in one
dimension.
Chaos, by itslef, does not lead to standard behaviour, since it does not
guarantee local thermodynamic equilibrium. A linear relation is found between
density fluctuations and temperature profiles. This link and the temporal
asymmetry of fluctuations of the main observables are robust against modifications
of thermostat parameters and against perturbations of the dynamics.
\end{abstract}

%\begin{keyword}
%Nonequilibrium fluctuations, reversibility, deterministic vs. stochastic dynamics
%
%\PACS
%05.40.-a, 45.50.-j, 05.45.-a, 02.50.Ey, 05.70.Ln
%
%\end{keyword}
%\thanks[corresponding]{Corresponding author. Telephone: +39+0522+522632, Fax: +39+0522+522616}
%\end{frontmatter}

%*******************************************************************

\section{Introduction}
Understanding the fluctuation properties of nonequilibrium phenomena is one of the major
tasks of modern statistical physics \cite{ESreview,RMM,BDGJL2007,BPRV,GGgenova,DenisReview,Gonnella,bsgj01,BDSJLcurrent}.
To this purpose, various generalizations of Onsager-Machlup theory
\cite{om53} have been proposed, like those of Refs.\cite{bsgj01,BDSJLcurrent}.
These papers have also
motivated various works on deterministic nonequilibrium particle systems
\cite{GR04,PSR06,GRV06,GRV07,PSR08} meant to investigate the relationship between
stochastic and deterministic models of nonequilibrium physics. One basic tenet of classical
statistical mechanics, indeed, maintains that the stochastic description is but a reduced
(mesoscopic) representation of the deterministic (microscopic) description of a given
system of interest. Nevertheless, the relation between the two representations is far from
fully understood. For instance, nonequilibrium steady states require thermostats, and
one expects the state of a macroscopic system to be unaffected by the details
of the thermostatting mechanism. Thus, even in mathematical models of nonequilibrium
steady states, one would like stochastic and deterministic thermostats to
lead to practically equivalent representations of the same phenomenon.\footnote{Rieder,
Lebowitz and Lieb, speaking of nonequilibrium
systems due to stochastic boundary reservoirs at different temperatures,
express this idea as follows: {\em ``the properties of a `long' metal bar should not depend
on whether its ends are in contact with water or with wine `heat reservoirs' at temperature
$T_1$ and $T_2$''} \cite{RLL}. In other words, the temperature should be the only thermodynamically
relevant property of a thermostat, in the nonequilibrium steady state.}
However, different theoretical models may capture different
aspects of a physical process, and a complete equivalence should not be
expected, except in some limiting situation, as postulated by various equivalence principles
\cite{ES93,CR98,TVB,GGequi,RS99,GRS,GG08}. The presence of local thermodynamic
equilibrium fosters the equivalence. Indeed, the thermodynamic relations are so
largely independent of the details of the microscopic dynamics, hence so generally valid,
precisely because of local equilibrium, cf.\ Section 4 below,
Refs.\cite{RMM,ESR07,LLP2,JR10,LMMP09,LMMP10} and references therein. In case local
equilibrium cannot be established, one may still wish to organize the different cases
in homogeneous groups, or universality classes \cite{CDP,LLP08}, to help our
understanding of nonequilibrium phenomena.

It is, then, interesting to point out the conditions under which the asymmetries of
fluctuations characterizing the (intrinsically irreversible) stochastic evolutions
\cite{bsgj01,lc97} are present in the dynamics of time reversal invariant particle systems.
These asymmetries may indeed be experimentally and numerically observed, and are thought
to be responsible for the irreversibility of macroscopic phenomena \cite{lc97}.
In Ref.~\cite{GR04}, the fluctuations of the current
of the nonequilibrium Lorentz gas, with large numbers $N$ of
noninteracting particles, were found
to be time-symmetric, despite the clearly irreversible behaviour of the system. As explained
in Ref.~\cite{GRV07}, this was due to the lack of interactions among the particles, hence to
the lack of correlations among them \cite{PSR08}, revealed by the large system
limit.\footnote{Asymmetric fluctuations are found at small $N$, since
the averages computed over small numbers of particles do not accurately
reproduce the correlation functions \cite{Germano}.}
On the other hand, the
absence of interactions prevents the onset of local thermodynamic equilibrium, despite the
convergence to a given steady state.
In Refs.~\cite{PSR06,GRV06,GRV07,PSR08}, it is therefore argued that systems of
properly interacting particles should typically have asymmetric fluctuation paths, as predicted for
stochastic systems. More precisely, Ref.\cite{GRV07} states
that temporally symmetric fluctuation paths of nonequilibrium time reversal invariant
deterministic dynamics require the untypical condition of vanishing correlations; generically,
fluctuation paths should then be asymmetric. Hence, deterministic reversible and stochastic interacting
particle systems are equivalent, i.e.\ belong to the same universality class, from
this point of view.

In the present paper, we continue the investigation of these issues, by looking at
variations of the nonequilibrium Fermi-Pasta-Ulam (FPU) model introduced by Lepri,
Livi and Politi
\cite{LLP1}. This model consists of $N$ anharmonic oscillators of
same mass $m$ and positions $x_j$, $j=1,..., N$, interacting via the nearest neighbours
FPU-$\zb$ potential \cite{FPU}:
\be
V(q) = k^2 \frac{q^2}{2} + \beta\frac{q^4}{4} ~,
\label{potential}
\ee
while the oscillators with $j=1$ and $j=N$ further interact with deterministic thermostats
known as Nos\`e-Hoover ``thermal baths'',
and with still walls. The internal energy of the system is then given by:
\be
H=\sum_{j=0}^N \frac 1 2 m \dot{q}_j^2+V(q_{j+1}-q_{j}) ~.
\label{energy}
\ee
The equations of motion take the form
\bea
&&m\ddot{q}_1 = F(q_1 - q_0) - F(q_2-q_1) - \xi_\ell \dot{q}_1 ~,
\nonumber \\
&&m\ddot{q}_j = F(q_j - q_{j-1}) - F(q_{j+1}-q_j) ~, \quad
\mbox{for } j = 2,..., N-1 \label{eqsmot} \\
&&m\ddot{q}_N = F(q_N - q_{N-1}) - F(q_{N+1}-q_N) - \xi_r \dot{q}_N, \nonumber
\eea
where $q_0 = q_{N+1} = 0$ represent the walls and $m$ is the mass of one particle, which we take to
be unitary, so that velocities $\dot{q}_i$ and momenta $p_i$ represent the same quantity.
Moreover, the interparticle forces are obtained differentiating the potential,
$F(q) = - V'(q)$, and the Nos\`e-Hoover thermostats, at ``temperatures'' $T_r$ and $\ts$ with
response times $\theta_r$ and $\theta_\ell$, are implemented by the variables
$\xi_\ell, \xi_r$, which obey
\be
\dot{\xi}_r = \frac{1}{\theta_r^2} \left( \frac{\dot{q}_N^2}{T_r} -1 \right)
~, \quad \quad
\dot{\xi}_\ell = \frac{1}{\theta_\ell^2} \left( \frac{\dot{q}_1^2}{\ts} -1
\right) ~.
\label{NHthermo}
\ee
The nonequilibrium FPU-$\zb$ model is {\em time reversal invariant},
but  {\em  dissipative}, in the sense that the time average of the divergence of the
equations of motion, $-(\langle \xi_r \rangle + \langle \xi_\ell \rangle)$, is negative.
Therefore, the sum of the Lyapunov exponents is negative, the phase space volumes
contract and the system approaches in time a nonequilibrium steady state, characterized by
a singular phase space probability distribution.

In the theoretical calculations, the motion of particles is commonly referred to the quantity $a$, which is the mechanical equilibrium distance between two nearest
neighbours \cite{LLP2}.
Then, the mean displacement of the position of particle $j$ from its
equilibrium position $ja$, is often assumed to be small, in order to define
microscopically the physical observables. It will be pointed out that this
is not appropriate, except for particularly small driving forces.

A quantity of interest is the ``local virial'' \cite{LLP1}. For particles in the bulk
($i=2,...,N-1$), this is the time average of the product of the displacement of each
particle, times the net force acting on it, $\langle q_{i} (F_{i+1}-F_i) \rangle$,
where we have set $F_i=F(q_i - q_{i-1})$ for simplicity. If the anharmonic part of $V$
is replaced by hard core elastic collisions, the interaction between particles is
purely harmonic, and one has $F_i-F_{i+1}= k^2(q_{i-1}-2q_i+q_{i+1})$. Therefore,
the local virial of particles in the bulk is expressed by
$$
k^2 \langle -q_i~(q_{i-1}-2q_i+q_{i+1}) \rangle ~, \qquad i = 2, ..., N-1
$$
while for $i=1,N$ it reads
$$
\langle q_1~(-\xi_L \dot{q}_1-k^2 (2q_1-q_2)) \rangle ~, \quad
\langle q_N~(-\xi_R \dot{q}_N-k^2 (2q_N-q_{N-1})) \rangle ~.
$$
An important feature of this model is that its long wavelength Fourier modes, representing the
slow relaxational dynamics, may be considered as approximately independent of the short
wavelength modes, representing the fast dynamics \cite{LLP2}, i.e.\ representing some sort of noise
added to the slow dynamics.\footnote{See, e.g.\ Ref.\cite{PPRV10} for
a discussion of the relation between fast and slow variables.} In the slow dynamics, one may
further separate a conservative non vanishing harmonic part, which cannot contribute to transport
phenomena, from a mode interaction part \cite{LLP2,SL98,DLLP}. From this standpoint, the
nonequilibrium FPU system is similar to the nonequilibrium Lorentz gas, as argued in
\cite{GRV07}, but the presence of nonvanishing correlations, even if decaying in time,
provides a mechanism for asymmetric fluctuation paths, within the corresponding decorrelation
time scales,
Refs.\cite{GRV06,GRV07}.\footnote{The role of correlations for temporal asymmetries has been
emphasized in Ref.~\cite{PSR08}, although in the different framework of homogeneously driven
and thermostatted systems.}

In this paper, we consider different values of $k^2$ and a different kind of anharmonicity:
that provided by hard core elastic collisions. We also consider the limiting cases in which
either the harmonic or the anharmonic interactions vanish. These limiting cases share indeed
some peculiarities of noninteracting particle systems, and constitute good candidates for the
study of the onset of temporally symmetric fluctuations \cite{GRV07,PSR08}. The fact
that all particles are identical, further means that there is no disorder in our chains, hence
no reasons for chaos, mixing or ergodicity, except those that may be contributed by the
boundaries. We will see that
nonequlibrium boundary conditions make correlations develop even in the bulk of such
particles chains, and lead to asymmetric fluctuations.
The results presented here lead to the following main conclusions:
\begin{itemize}
\item[{\bf a)}]
The kinetic temperature profiles are linearly related to the deviations of
the mean positions
of the particles from their equilibrium values, hence to the deviations
from uniformity of the density profiles (Section 2):
\be
T_i=\beta_1 \langle x_{i+1}-x_{i}\rangle+\beta_2 ~.
\label{LinRel}
\ee
This relation and the temporal asymmetry of fluctuations are robust
against modifications of the parameters of the thermostats and of the interaction potentials,
besides being robust against stochastic perturbations of the deterministic dynamics (Appendix B).
\item[{\bf b)}] Comparison between our results for deterministic, time reversal invariant,
dynamics and those found in the literature for chains with stochastic thermostats, shows
that the behaviours concerning the two kinds of thermostats are only qualitatively similar.
The two kinds of thermostats do not belong to the same universality class, except in a
coarse sense; equivalence is then only expected in the equilibrium limit.
\end{itemize}
Other relevant conclusions include:
\begin{itemize}
\item[{\bf c)}]
The deviations of the mean positions of the particles from their equilibrium values
are large, except in the purely harmonic case which, however, does not sustain a
temperature gradient. Hence the microscopic definitions of nonequilibrium thermodynamic
observables, based on small deviations, are generally impaired (Section 2).
\item[{\bf d)}]
As in stochastically thermostatted systems, investigated by other authors, the properties of
steady states depend substantially on the microscopic details of the dynamics, showing that
the absence of genuine local thermodynamic equilibrium is typical of physics in 1-d
(Section 2).
\item[{\bf e)}]
As in the stochastic case, temporal asymmetries are ubiquitous in deterministic,
time reversal invariant,
nonequilibrium particle systems, as long as correlations among particles are present.
Such asymmetries are not restricted to large deviations
in the large system limit and do not require local thermodynamic equilibrium (Section 3).
\item[{\bf f)}] At variance with common expectations, Nos\'e-Hoover
thermostats alter the bulk behaviour, even when the bulk resembles a noninteracting particle
system, for both small and large temperature gradients. This is a
manifestation of the nonlocality generically observed in nonequilibrium steady states
\cite{bsgj01,DLS,DKS,HSpohn} (Section 4).
\item[{\bf g)}] Chaos by itself is not sufficient to establish standard behaviour in 1-d
systems, besides not even being necessary \cite{LLP1,JR06,Squares,LLP97} (Section 5).
\end{itemize}

In Section 2, we consider hard spheres on a line, connected in pairs by harmonic
forces. In Section 3, the temporal symmetries of the fluctuations of this model are
analyzed. In Section 4, the limiting cases of purely harmonic forces and
purely elastic collisions are considered. In section 5, the role of chaos is
investigated. Section 6 is devoted to concluding remarks. Appendix A quantifies
the deviations of the temperature profiles from a theoretical curve. Appendix B
gives results about stochatstically perturbed dynamics.

\newpage

\section{Hard core and harmonic potentials}
In the present case, the harmonic part of the potential of Eq.(\ref{potential}) is retained, while
the term proportional to $\zb$ is replaced by hard core elastic collisions between particles of
radius $r=1$. This model shares some features of non-interacting particles systems. Indeed, isolated
one-dimensional systems of elastic particles of equal mass and size exchange their momenta in such a
way that the overall motion is equivalent to that of a system of non-interacting particles, in which
each particle preserves its momentum \cite{Wojtk}. Moreover, in isolated systems, the harmonic potential
should not constitute an important source of correlations, since the oscillations it entails
amount to independent normal modes. Nevertheless, two features distinguish our model from
systems of non-interacting particles:
\begin{itemize}
\item the elongation of the harmonic springs discriminates the case in which
particles bounce back at collisions, from the case in which they pass through one another;
\item the presence of the thermostats at the boundaries of the chain spoils the
normal modes decomposition of the purely harmonic motion.
\end{itemize}
The question, which we have investigated numerically, is how crucially these facts affect
the behaviour of the system. To verify that the steady state is indeed achieved, we have
computed the local ``temperatures'',
$\langle p_i^2 \rangle$, and the virial profiles, at different times along a simulation of length $t_{_{\mbox{max}}}$. We have found that the temperature profile rapidly approaches
its asymptotic form (a quarter of our $t_{_{\mbox{max}}}$ suffices for good accuracy),
while the virial expression takes much longer to settle down, cf.\ Figure 1 for the case
with $N=400$, $k^2=1$, $\theta_\ell=\theta_r=1$.

Similar profiles have been found in numerous one-dimensional
nonequilibrium chains of oscillators, with anharmonic interaction potentials, like the
FPU-$\zb$ chains, \cite{LLP2,Aoki,DharAP}. In particular, assuming that dynamical chaos practically
amounts to some degree of stochasticity in the evolution, our model should behave similarly
to that of Ref.\cite{LMMP09}, which has stochastic boundary thermostats, harmonic interactions
and stochastic exchanges of momenta between nearest neighbours, meant to mimic (momentum and
energy preserving) hard core interactions. In a suitable continuum limit, the corresponding
energy profile takes the form \cite{LMMP09}:
\be
T(x) = \frac{T_\ell+T_r}{2} + \frac{\sqrt{2} \left( T_\ell - T_r \right)}
{\left( \sqrt{8} - 1 \right) \zeta{(3/2)}} \sum_{n=0}^\infty
(2n+1)^{-3/2} \cos \left( \frac{(2n+1) \pi}{2} (x+1) \right)
\label{TOFX}
\ee
where $\zeta$ is the Riemann $\zeta$-function, and the space variable
is rescaled so that the left end of the chain corresponds to $x=-1$ while the right end
corresponds to $x=1$.

However, as shown by the left panel of Figure 2, and deduced from a comparison of our Figure 1
with Figure 2 of \cite{LMMP09}, the temperature profiles obtained by Lepri, Mej\'ia-Monasterio
and Politi
are not completely equivalent to ours. In the first place, the slope of profile (\ref{TOFX}) has
square root divergences at the boundaries, which are less steep than ours. Secondly, the profiles
of Ref.\cite{LMMP09} are odd with respect to the centre of the chain, even at finite $N$, while
our profiles are not. One further
difference is the simple $k^2$ dependence of the finite$-N$ temperature profiles of
\cite{LMMP09}, as opposed to the strongly irregular dependence of our profiles, cf.\
central panel of Figure 2. To investigate this question, we have analyzed the
case with $k^2=1$, $T_\ell=320$, $T_r=20$, $\theta_\ell = \theta_r = 1$, for $N = 100, 200, 250, 300, 400$.
Differently from \cite{LMMP09}, we found no evidence of the $N^{-1/3}$ rate of convergence towards the
asymptotic profile $T(x)$ and, in most of the chain, the difference between our profiles and $T(x)$
was observed to grow with $N$. Therefore, as shown in Appendix A, either a scaling regime towards
$T(x)$ sets in at $N$ larger than 400, or our asymptotic profiles will not correspond to those of
\cite{LMMP09}. On the other hand, Appendix A shows that our profiles converge rapidly to
a given asymptotic profile, which is not odd. This may be related
to the fact that, differently from the stochastic case, the energy exchange between Nos\'e-Hoover
boundary thermostats and 1-d (or almost 1-d) systems of particles is more efficient at the hot side
than at the cold side, cf.\ Ref.\cite{MMR08} and Section \ref{Harmonic} below.\footnote{In
Ref.\cite{MMR08} it was observed that small temperature gradients allow the cold thermostat
to behave as ``thermodynamically'' as the hot thermostat. The boundary thermostats of 1-d
chains act only on momentum degrees of freedom, making it
difficult, especially at large temperature gradients, for local equipartition to be
established, except in the presence of mass diffusion \cite{Carlos}. However, the difference
between $T$ and the profiles obtained by Nos\'e-Hoover boundary thermostats persists
at small gradients, as further evidenced in Section \ref{Harmonic}.}

\vskip 15pt

\centerline{\includegraphics[width=6.8cm,height=5.5cm]{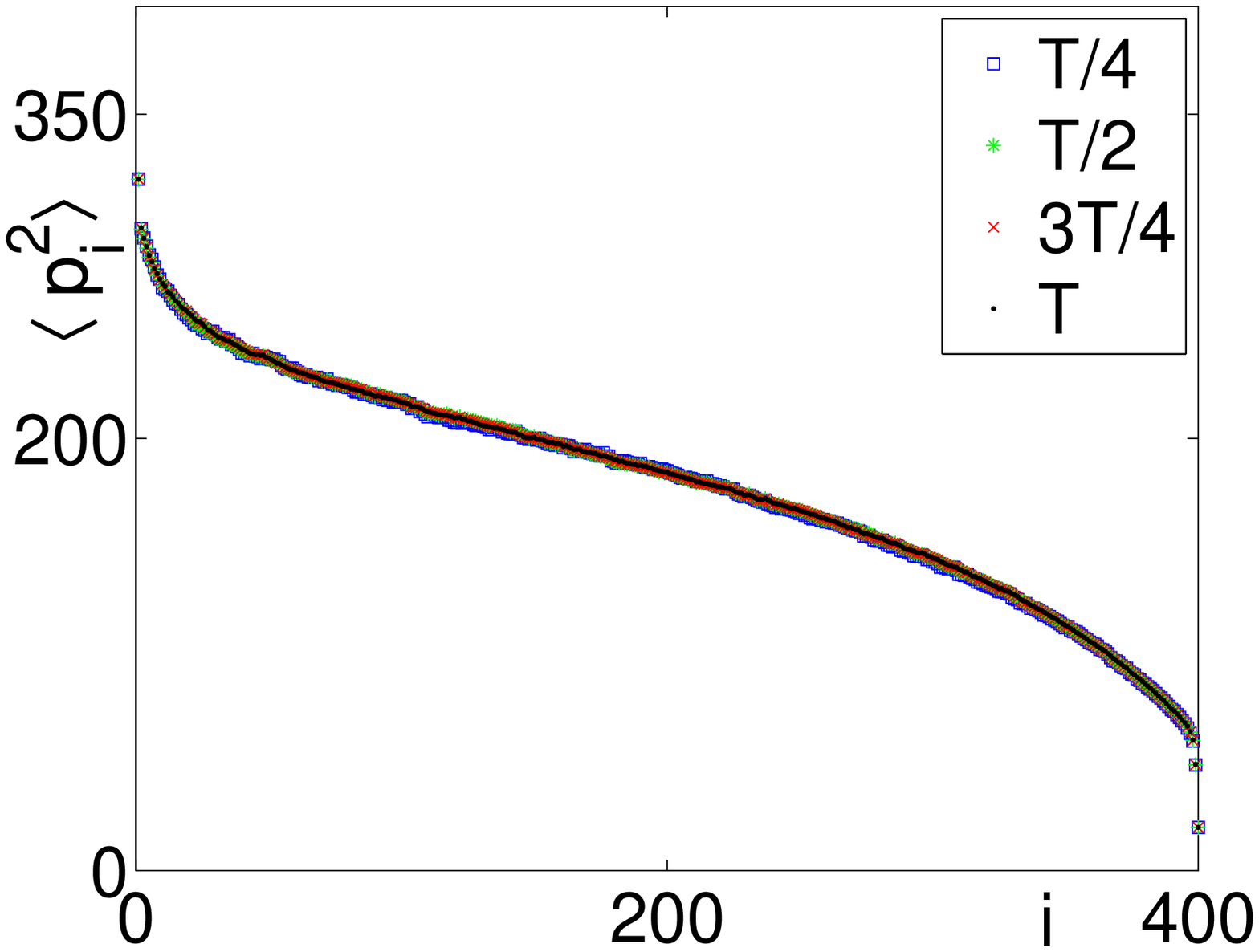} ~~~
\includegraphics[width=6.8cm,height=5.5cm]{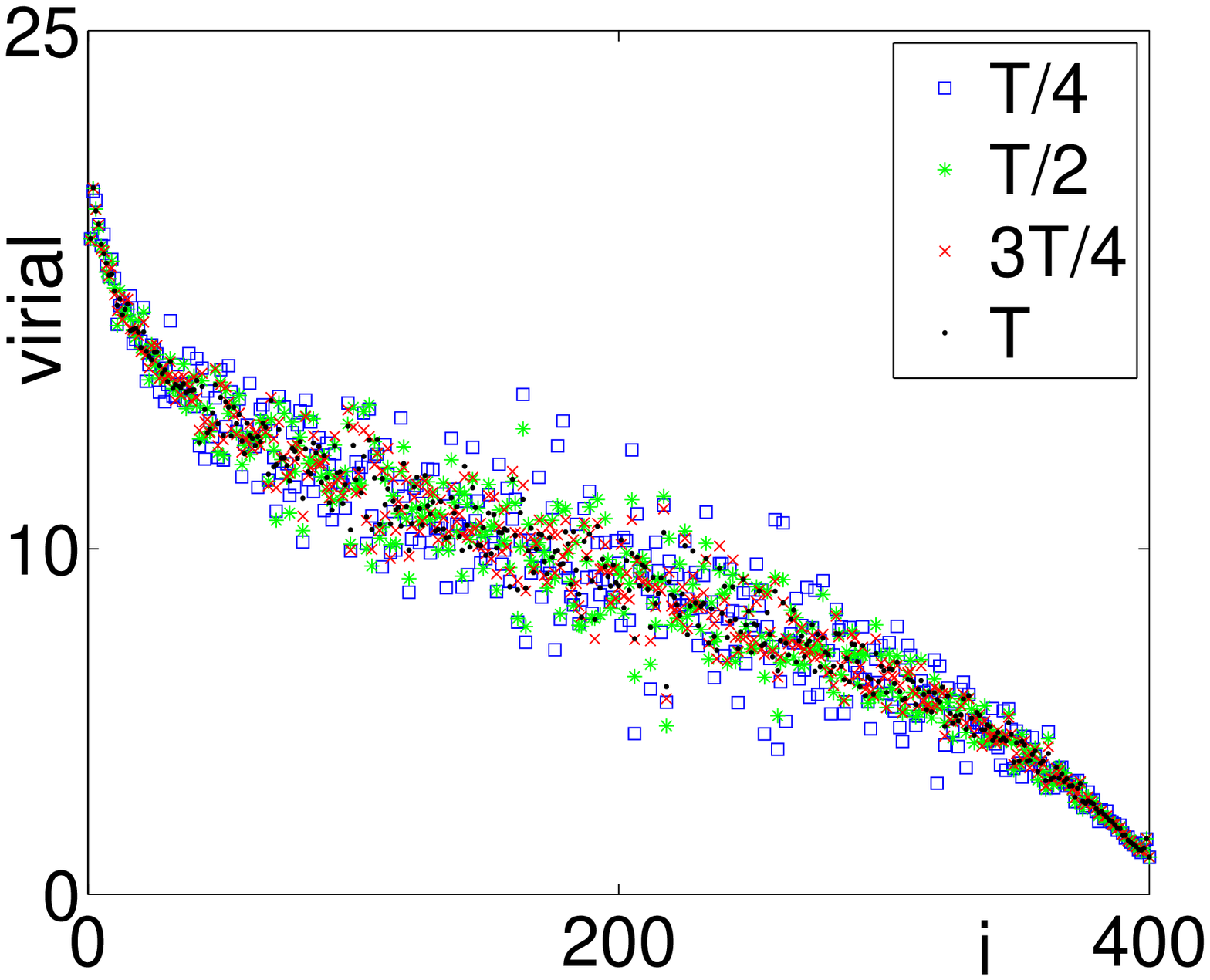}}
\vskip 0pt
{\small {\bf Figure 1.} (Color
 online) Kinetic temperature (left panel) and local virial (right panel)
profiles for different fractions ($1/4$, $1/2$, $3/4$, 1) of the simulation length
$t_{\mbox{max}} = 2.5 \cdot 10^5$ time steps of size $h=10^{-4}$, with $N=400$, $k^2=1$,
$\theta_\ell=\theta_r=1$. The $\langle p_i^2 \rangle$ profile reaches rather rapidly
its asymptotic form (the curves for the four simulation lengths are indistinguishable),
while the virial expression takes longer to converge. Because of hard core interactions,
kinetic temperature and virial do not coincide \cite{TKS}.}

\vskip 15pt

Despite these facts, the temperature profiles concerning nonequilibrium chains of oscillators
typically enjoy the same qualitative behaviour, consisting of steep curves at the boundaries,
due to contact resistance, interpolated by almost straight lines in the bulk. The nature of
this behaviour is far from obvious, cf.\ the discussion in Ref.\cite{LMMP09,DharAP}. In any
event, the strong
and irregular dependence of the steady states of both deterministic and stochastic 1-d systems,
on microscopic details of the dynamics, as well as the violations of Fourier Law,
reveal the absence of genuine local thermodynamic equilibrium and of diffusion,
see e.g.\ Refs.\cite{RLL,LLP2,JR10,LMMP10,Aoki}.

For certain observables, this dependence on the microscopic mechanisms, hence the lack
of local thermodynamic equilibrium, persists in the large $N$ limit, \cite{RLL,LMMP09,LMMP10}.
Similarly, completely different situations are determined by the boundary conditions, such as
the divergence of the conductivity of disordered harmonic chains with free boundaries, in the
thermodynamic limit, as opposed to its vanishing trend in chains with fixed boundaries
\cite{LMMP10,LLP08}. Moreover, the scaling behavior of the
conductivity with system size crucially depends on the spectral properties of the
heat baths \cite{dhar}.

\vskip 15pt

\centerline{
\includegraphics[width=5.8cm,height=5.5cm]{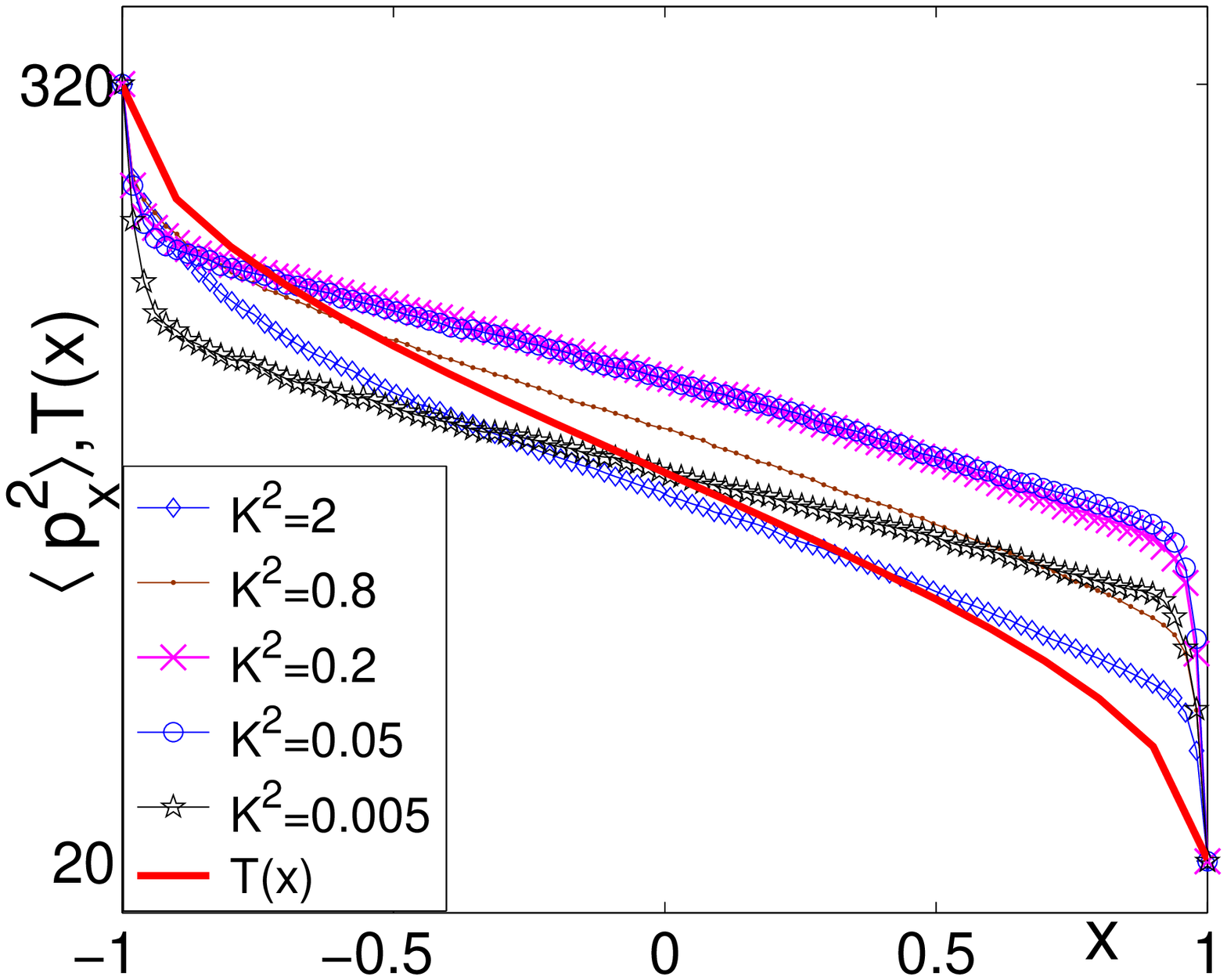}
\includegraphics[width=5.5cm,height=5.5cm]{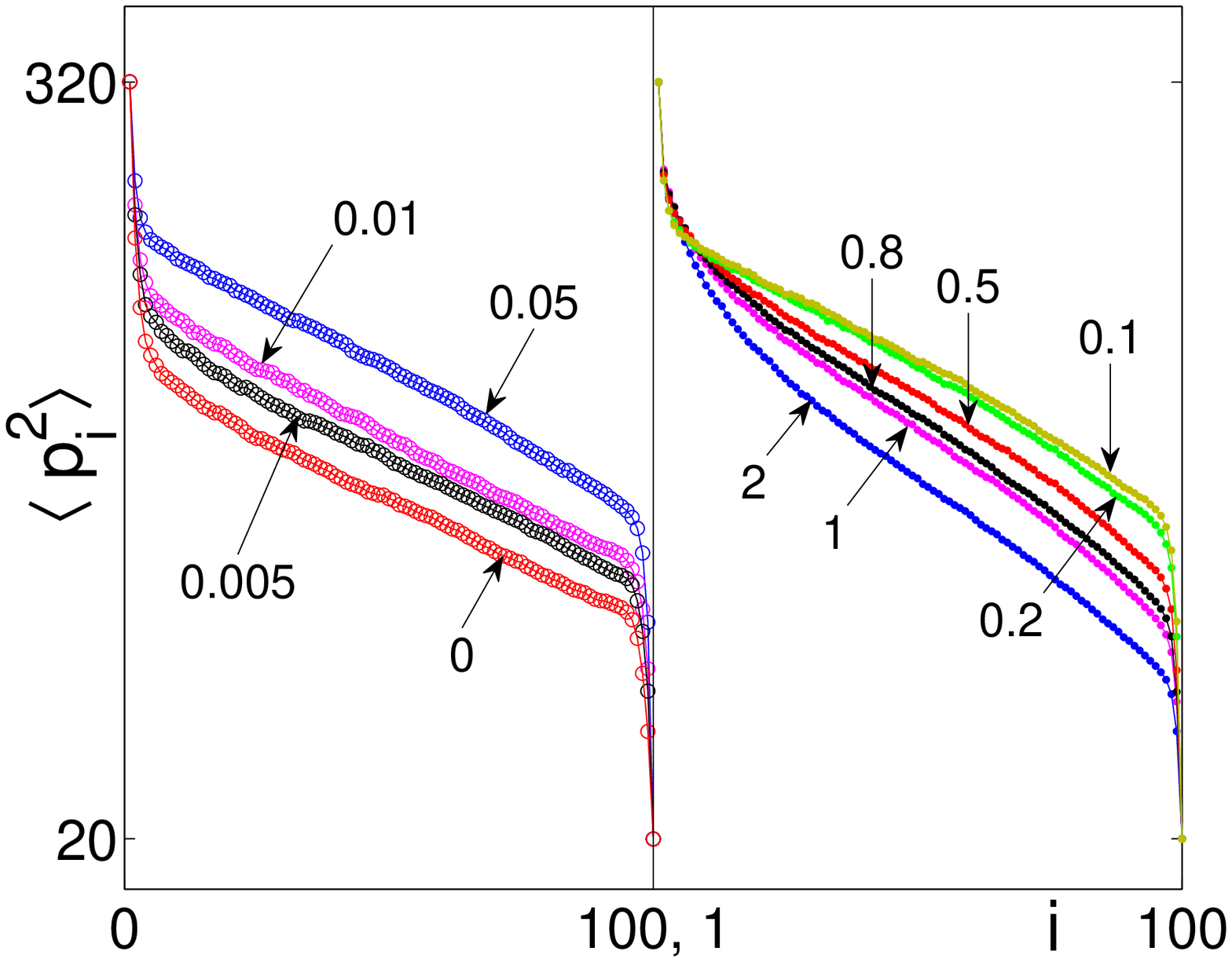}
\includegraphics[width=5.8cm,height=5.5cm]{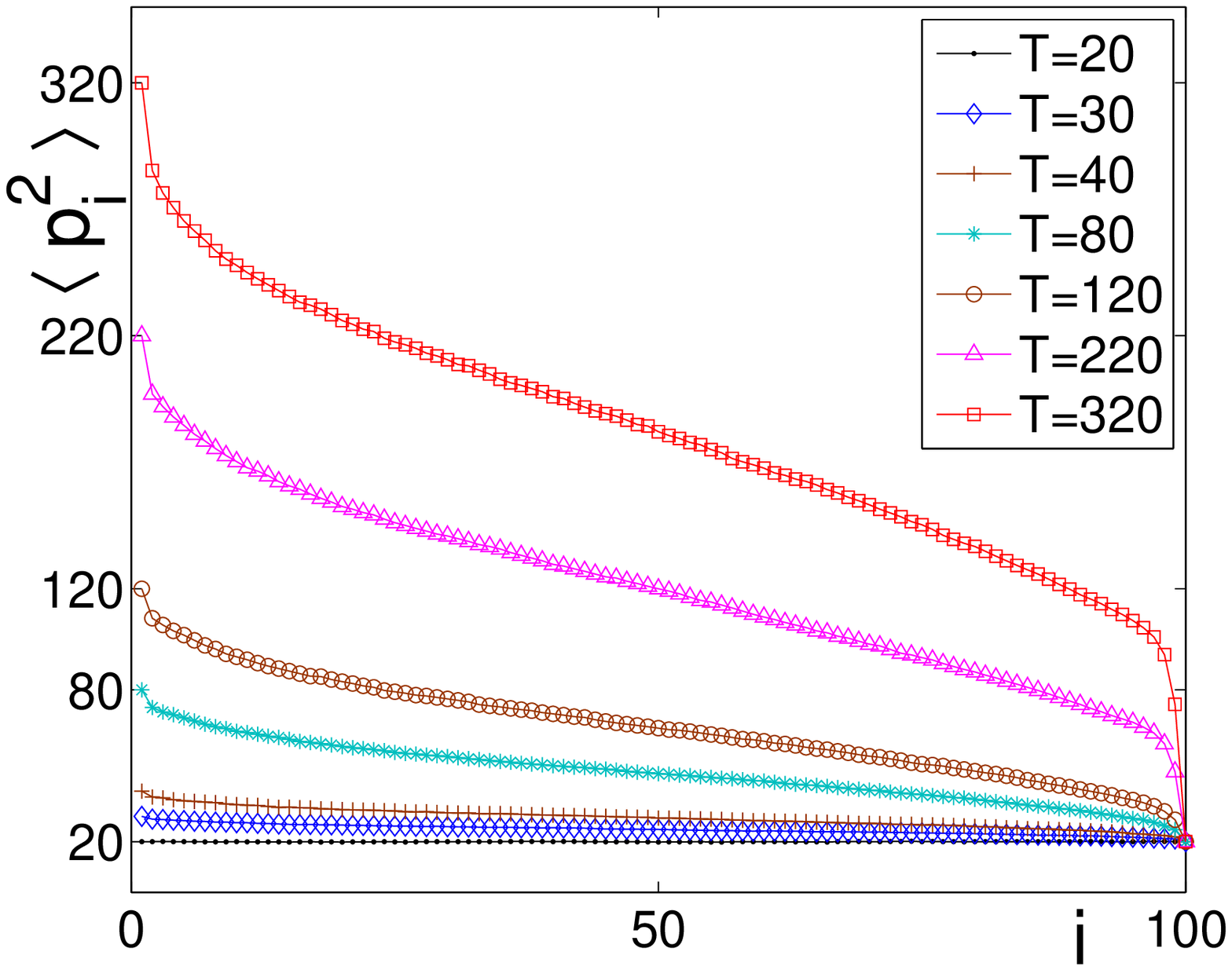}
}
\vskip 2pt
{\small {\bf Figure 2.} (Color
 online) Comparison between the temperature profile $T(x)$, eq. (\ref{TOFX}), of Ref.\cite{LMMP09} (continuous
line) and those of our deterministic models, for $N=100$, $T_\ell=320$, $T_r=20$ and decreasing $k^2$
(left panel). Dependence of the deterministic profile on $k^2$ (central panel). For $k^2 \to 0$, it
tends to the $k^2=0-$profile in an irregular fashion. Arrows indicate the values of $k^2$.
Deterministic profiles as functions of $T_\ell$, with $k^2=0.1$ and $T_r=20$ (right panel).}

\vskip 15pt

Rieder, Lebowitz and Lieb interpreted this kind of results as {\em ``unphysical''} \cite{RLL}.
Indeed, were 1-d chains of oscillators to represent macroscopic objects enjoying thermodynamic
properties, the thermostat dependence of the heat fluxes of \cite{RLL}, or of our temperature
profiles, would be unrealistic. However, recent theoretical and technological
developments allow different interpretations. The study of nearly 1-d systems, especially
with non-macroscopic numbers of elementary constituents, indicates that standard
thermodynamic relations typically fail to describe their behaviour,
although a complete understanding of their properties is currently missing, see e.g.\
\cite{LLP08,DharAP,NatureNano,Chang}
and references therein. Thus, a possible absence of local
thermodynamic equilibrium does not need to be unphysical. Indeed,
local thermodynamic equilibrium requires a sufficiently
fast decay of correlations of all relevant observables, which would lead to diffusive transport.
But this is seldom afforded by (quasi-) 1-d systems
\cite{JR10,JBR08},\footnote{For instance, certain correlations never decay in the narrow
channels of Ref.\cite{JBR08} and in our chains of hard core particles, since the particles
order is preserved in time. While this is not a problem {\em per se} --order is preserved
in 3-d crystals as well-- it violates the hypothesis of molecular chaos and bears
substantial consequences on transport phenomena.} which include
modern technological artifacts as well as structures found in nature.
It is also known that 1-d systems are affected by peculiarities such as the cumulative
$O(N)$ fluctuations, about the particles equilibrium positions \cite{Peierls}. These
frustrate the microscopic definitions of observables based on the assumption
of small fluctuations and, more importantly, are at odds with the properties of
solids. Furthermore, we have found that temperature gradients induce large displacements
of the average equilibrium positions, making the distribution of matter inhomogeneus.
For three different values of $k^2$, the left panel of Figure 3 portrays these average
deviations, $\langle q_i \rangle = \langle x_i - ai \rangle$, for $i=1,...,N=100$.
The right panel of the figure portrays the quantity $\langle x_{i+1} - x_i \rangle$,
which is related to the inverse of the density of particles.
One observes that nonequilibrium boundary conditions lead to a shift of all particles
towards the cold side of the chain, hence to a density gradient.

Denoting by $\langle \cdot \rangle_{k^2}$ the averages computed with elastic constant
equal to $k^2$, closer examination shows that:
\begin{itemize}
\item
all $\langle q_i \rangle_{k^2}$ are positive and, in the bulk,
are approximated by a parabola with maximum at ${i} > N/2$ for $k^2 \ne 0$,
while the maximum of $\langle q_i \rangle_0$ occurs at ${i} = N/2$;
\item
consistently,
$\langle q_N \rangle_{k^2} > \langle q_1 \rangle_{k^2}$ for $k^2 \ne 0$, while
$\langle q_N \rangle_{0} = \langle q_1 \rangle_{0}$;
\item
$\langle q_N \rangle_{k^2} = \langle q_N \rangle_{\tilde{k}^2}$ for
all pairs $k^2 , \tilde{k}^2$, while
$\langle q_1 \rangle_{k^2} = \langle q_1 \rangle_{\tilde{k}^2}$ only if both $k^2 , \tilde{k}^2 \ne 0$;
\item
the dependence of $\langle q_i \rangle_{k^2}$ on $k^2$ is not monotonic,
like the temperature profile is not.
\end{itemize}

Therefore, the systems with pure hard sphere interactions, whose bulk corresponds to
noninteracting particles, behave differently from the systems with both hard core and
harmonic interactions. Indeed, while the temperature profiles at $k^2=0$ do not substantially differ from those at $k^2 \ne 0$, the net energy transfer with $k^2=0$ is singular,
since it is equivalent to that of a single particle, bouncing back and forth between the
hot and cold walls, independently of the value of $N$.

%\newpage

\vskip 15pt

\centerline{
\includegraphics[width=6cm]{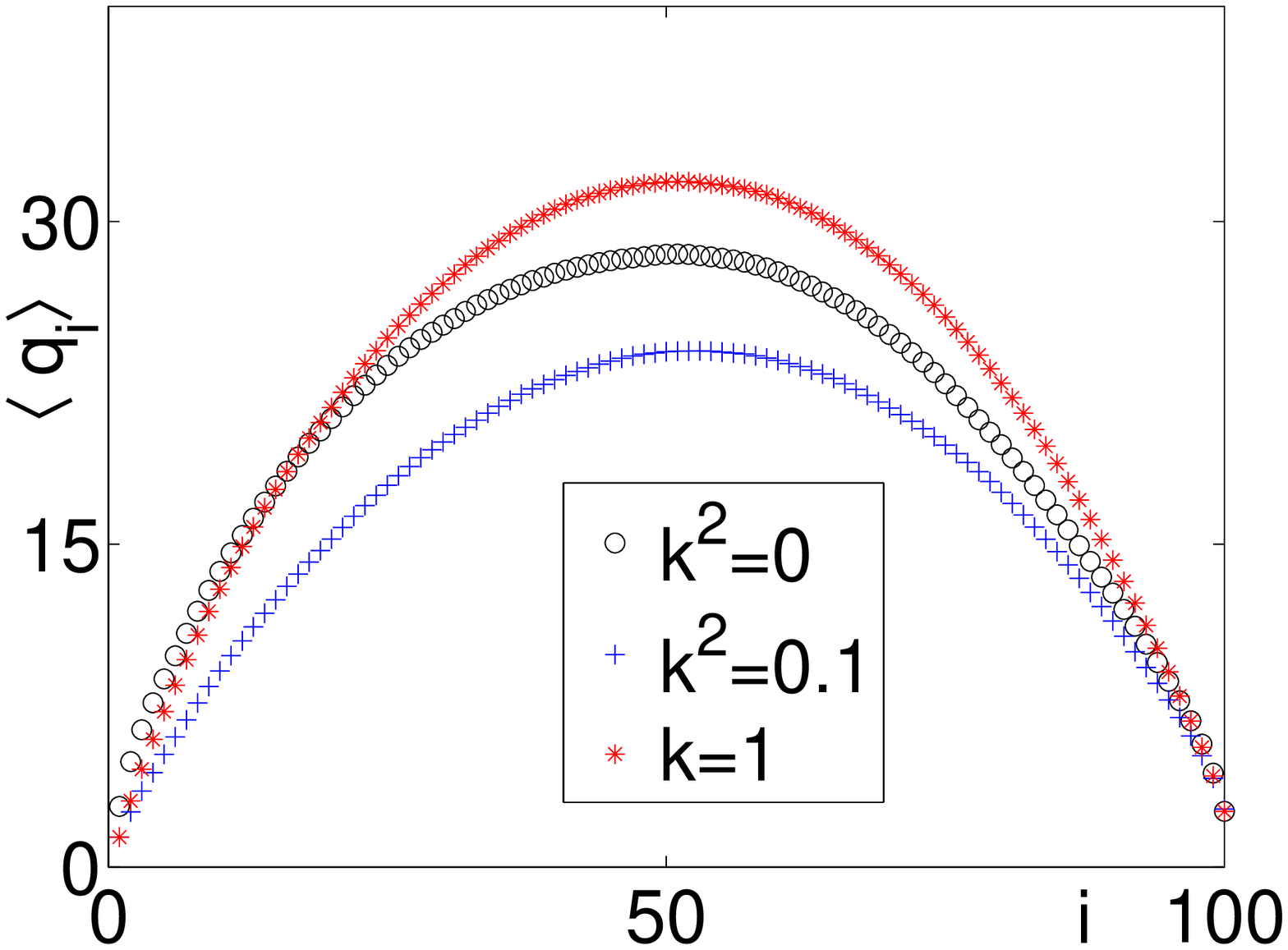} ~
\includegraphics[width=6cm]{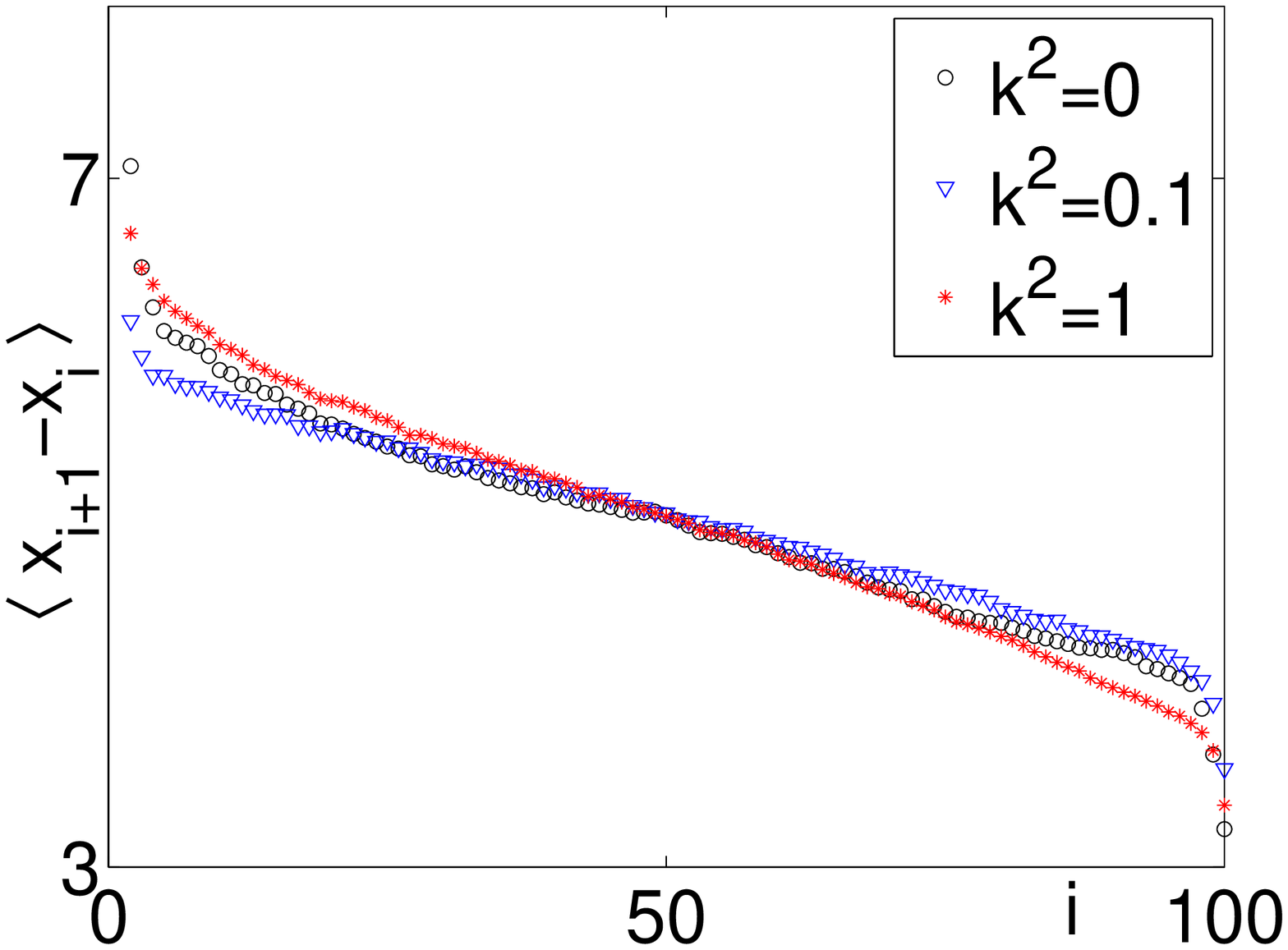} ~
}
\vskip 3pt \noi
{\small {\bf Figure 3.} (Color
 online) Mean displacement $\langle q_i \rangle_{k^2}$
from the equilibrium centres of oscillation of the particles, $ai$ (left panel).
Mean coordinates difference
$\langle x_{i+1} - x_i \rangle$, for several vales of $k^2$ (right panel).
%(central panel).
The higher this difference, the lower the particles density. In both panels $N=100$, $T_\ell=320$, $T_r=20$.
%Temperature profiles obtained from
%integration of Eqs.(\ref{Flaw},\ref{kappa}) with different values of $q_\ell$, $q_r$
%and $\za$, for $T_\ell=320$ and $T_r=20$ (right panel). In the
%left part of this plot, profiles decrease as $\za$ increases, taking values
%$-1, 1/6, 1/2, 2/5, 1$.
}

\vskip 15pt

Recalling that $a=5$, in our calculations, these plots show that the steady state
(average) particle distribution is not uniform and that the deviations of
particles positions from their equilibrium values are large and correlated to the
kinetic temperature profiles. Indeed,
$\langle x_{i+1}-x_i \rangle$ which, apart from an unessential additive constant,
is the discretized derivative of $\langle q_i\rangle$, qualitatively
approximates the
typical temperature field (compare e.g.\ the right
%cental
panel of Figure 3
with Figures 1 and 2),
hence a linear relation between the two quantities may be surmised:
\be
T_i=\beta_1 \langle x_{i+1}-x_{i}\rangle+\beta_2 ~.
\ee
%This fact suggests a relation between thermal conductivity $\kappa$ and
%distribution of $\langle q_i \rangle$, if Fourier's law is assumed to hold.
%Indeed, in the steady state, one has:
%\be
%\frac{\rm d}{{\rm d} x} \left[ \kappa(x) \frac{{\rm d} T}{{\rm d} x} \right] = 0 ~,
%\quad \mbox{i.e.} \quad  \frac{{\rm d} T}{{\rm d} x} = \frac{\rm C_1}{\kappa(x)}
%\label{Flaw}
%\ee
%where $C_1$ is an integration constant, and, assuming a connection
%between $T$ and $\langle q_i\rangle$, one obtains a relation between $k$
%and $\langle q_i\rangle$ as well.
%Because the $\langle q_i \rangle$ are well approximated by a parabola in most
%of the chain,\footnote{This is
%shown by the almost-linear behavior of $\langle x_{i+1}-x_{i}\rangle$, which
%is only violated at the borders of the chains.} let us investigate the consequences of taking
%\be
%\kappa(x) \propto \left[ (x-q_\ell)(N+q_r-x) \right]^\za
%\label{kappa}
%\ee
%where $q_\ell,q_r$ are positive small constants, and $\za \in \zR$.
%
%A comparison between the temperature profiles of the right panel of Fig.3 and
%those obtained numerically shows that they are, indeed, qualitatively similar,
%confirming that $\kappa$ is correlated with the displacements $\langle q_i \rangle$.
%%The good agreement between $T_i$ thus defined and the derivative of
%%$\langle q_i\rangle$ is demonstrated by Fig.4.

The good agreement between $T_i$ and the kinetic temperature profile  $\langle p_i^2\rangle$ is demonstrated by Fig.4.

\vskip 15pt

\centerline{
\includegraphics[width=6cm]{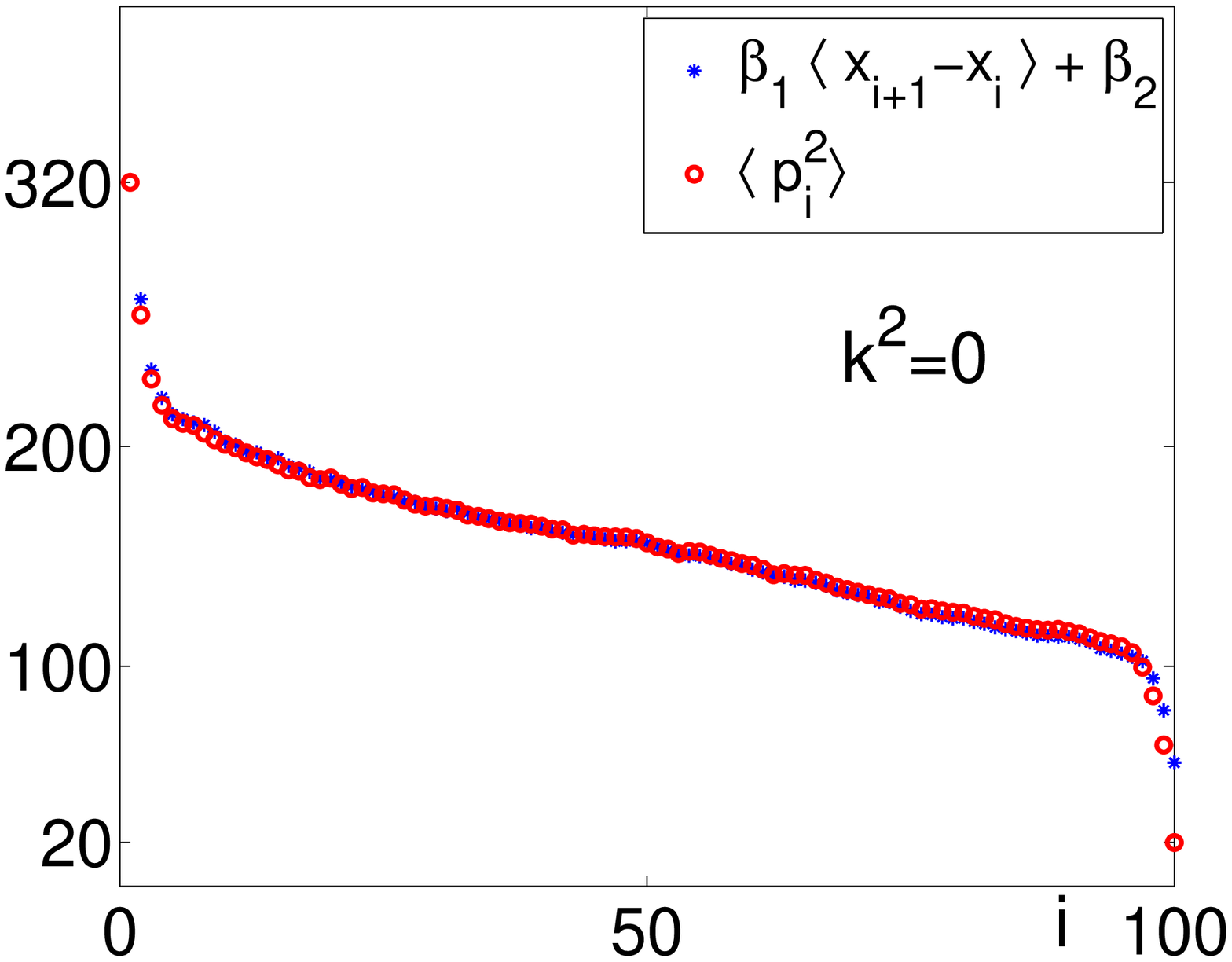} ~
\includegraphics[width=6cm]{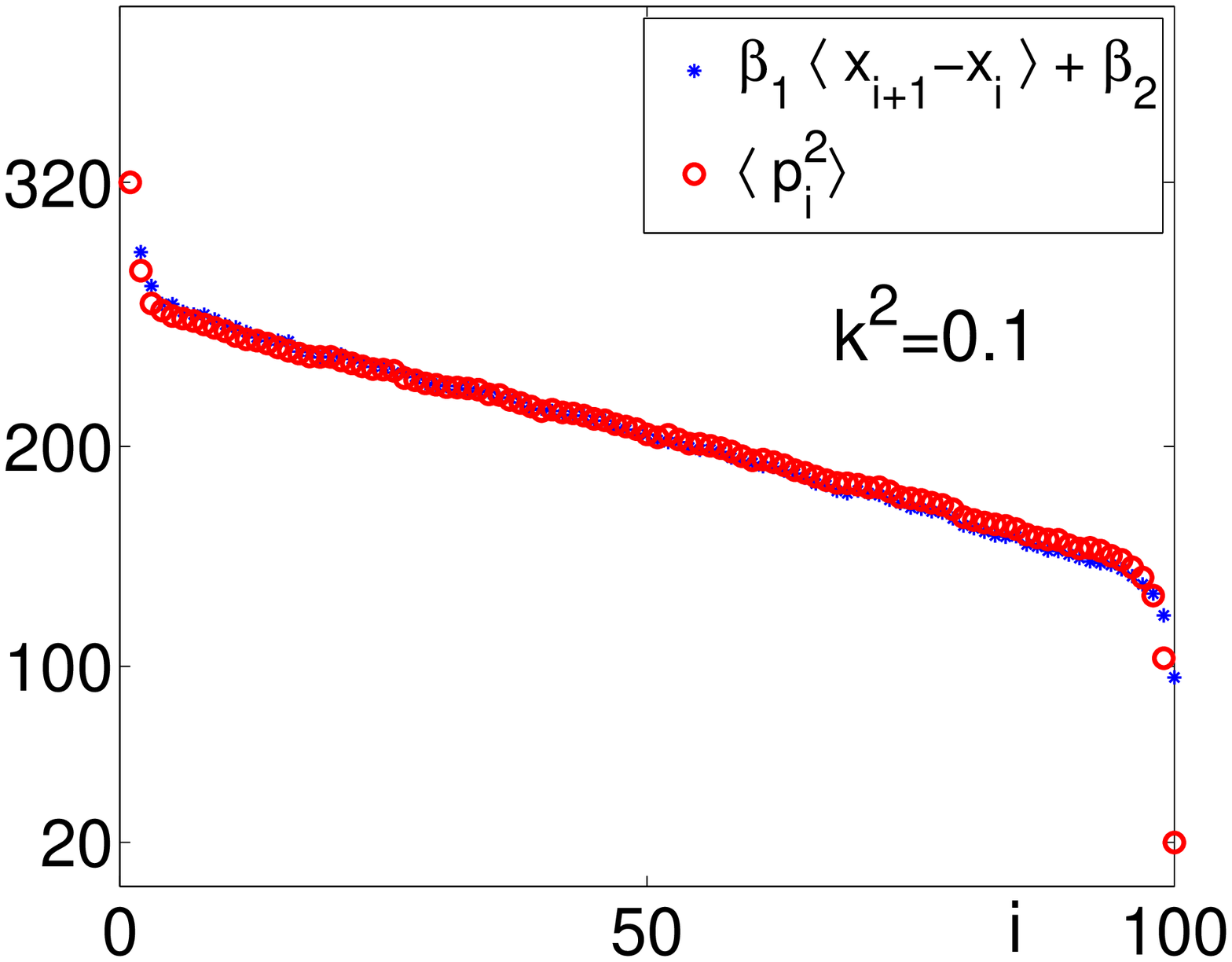} ~
\includegraphics[width=6cm]{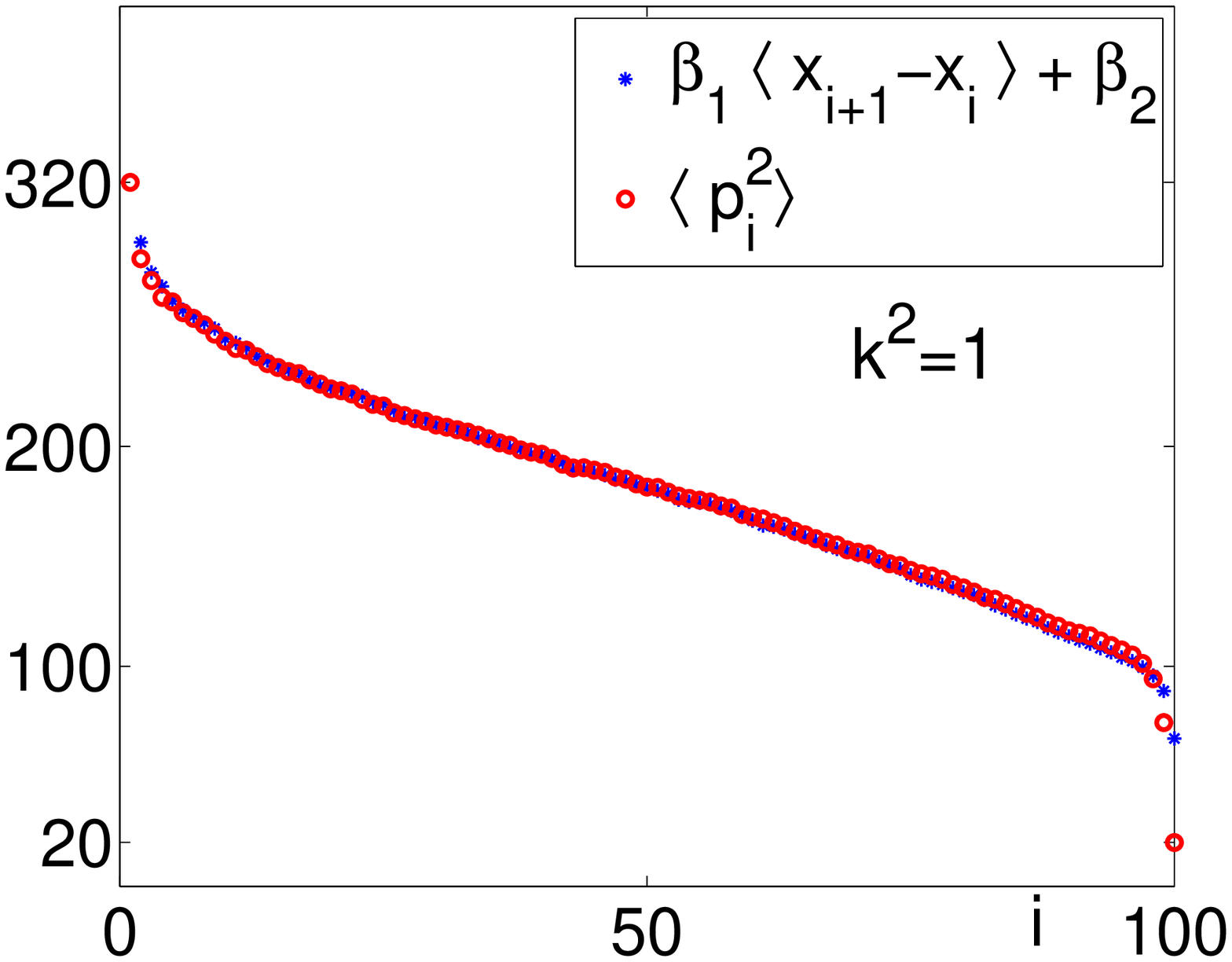}}
\vskip 3pt \noi
{\small {\bf Figure 4.} (Color
 online) Comparison between temperature profiles (red circles) and rescaled average displacements of neighbouring particles $\beta_1\langle x_{i+1}-x_i \rangle +\beta_2$ (blue stars), for $N=100$, $T_\ell=320$, $T_r=20$.
The constants $\beta_1,\beta_2$ are obtained by a least-square fit of the data of
Fig.3. The values of the elastic constant are: $k^2=0$ (left panel), $k^2=0.1$ (central panel), $k^2=1$ (right panel).}

\vskip 15pt
%
%Becasue the precise shape of the profiles obtained from Eqs.(\ref{Flaw},\ref{kappa})
%depends practically only on $\za$, a different light may be shed on the problem of the
%divergence with $N$ of the mean conductivity $\overline{\kappa}$,
%since our assumptions lead to
%\be
%\overline{\kappa} = \frac{1}{N} \int_0^N \kappa(x) ~ \mbox{d} x \sim N^{2 \za}
%\label{meankappa}
%\ee
%Thus one could conjecture that, in 1-d chains of oscillators, with harmonic and hard
%core interactions, $\overline{\kappa}$ diverges as $N^{2 \za}$, where $\za$ is obtained
%by fitting the solution of Eqs.(\ref{Flaw},\ref{kappa}) to the temperature profiles.
%As other chains of oscillators lead to similar profiles, this hypothesis
%could be more widely applicable.
\noi
This relation between temperature profiles
and rescaled average displacements of neighbouring particles is robust against
modifications of all the parameters of the dynamics, including thermostat parameters
and interaction potentials, besides being robust against (small and large) stochastic
perturbations of the dynamics, cf.\ Appendix B. Apart from the temporal asymmetries of
the fluctuations of the main observables, considered in the next section, this is the
only result which does not show a delicate dependence on the details of the microscopic
dynamics.

\newpage

\section{Temporal symmetry of fluctuation paths}

In the literature, various notions of fluctuation path have been investigated. Given the
substantial equivalence of the results based on such different notions, we adopt the first
definition of fluctuation-relaxation path of Ref.\cite{GRV07}, denoted by FR1.\footnote{See \cite{GRV06,GRV07} for a discussion of the
ambiguities in the definitions of fluctuation paths in deterministic dynamics. Despite
such ambiguities, the different notions of fluctuation path have led to analogous conclusions
and can be used interchangeably to asses the symmetry properties of nonequilibrium fluctuations.
Furthermore, the different notions coincide in the large $N$ limit, cf.\ \cite{PSR06,GRV06,GRV07,PSR08}.}
Denote by $\mathcal{M}$ the phase space of our system, by $S^t : \mathcal{M} \to \mathcal{M}$
the time evolution operator, and by $X: \mathcal{M} \to \zR$
an observable of interest. Consider an initial phase $\zG \in \mathcal{M}$, in the support of
the steady state phase space probability distribution, and denote by $\xt$ the quantity
$X(S^t \zG)$, under the assumption that all, but a set of vanishing probability, such initial
conditions enjoy the same statistics. Choose a fluctuation value $\mathcal{T}(X)$.
The FR1 fluctuation path is defined as follows:
\vskip 5pt
\noi
{\bf Definition. }{\it Assume that $X_{\hat{t}}=\mathcal{T}(X)$. Then, for any $t_0 , \zt > 0$,
the FR1 fluctuation path of duration 2$t_0$ based at $\hat{t}$ is the curve
$$
\{ X_{\hat{t}+\tau}: \tau \in [-t_0,t_0] \}
$$
in the $(\zt,X)$-plane.
}

\vskip 5pt
\noi
As in Ref.\cite{GRV07}, the symmetry properties of the fluctuation paths of the observable $X$
are here assessed setting a threshold $\mathcal{T}(X) = \zE(X) + 3\zs(X)$, three standard deviations
above the mean $\zE(X)$, while the width of the observation time interval
is taken to be $2 t_0 = 2$. Then, all such time intervals are translated by $-\hat{t}$, so that the centers of all fluctuations coincide at $t=0$.
%It appears that relaxation times, hence sampling times, increase with $N$.

The first fluctuating observable we consider is the local dimensionless density
of particles in the center of the chain, defined in a box of size $L$,
ideally with $1 \ll L \ll N$:
\be
\rho={\rho_0}^{-1} \frac{L}{x_F-x_I},\quad x_I=\min_{j\in {\cL}} x_j,\quad
x_F=\max_{j\in {\cL}} x_j ~,
\ee
where $\rho_0=L/(L-1)a$ is the equilibrium density.

For a family of $n$ (time translated) fluctuations,
$\left\{\rho^{(s)}(t), t\in [-t_0,t_0]\right\}_{s=1}^n$, we obtained
the average path $\overline{\rho}_\tau$ subdividing $[-t_0,t_0]$ in
$b_0=100$ bins, and by computing
\be
\overline{\rho}_\tau=\frac{1}{n\, m(\tau)}\sum_{s=1}^n \sum_{t=-t_0}^{t_0} \rho^s(t)\mathcal{C}_\tau(t)
\ee
where $\tau$ labels a bin, and $\mathcal{C}_\tau(t)=1$ if $t$ belongs to the $\tau$-th bin, while $\mathcal{C}_\tau(t)=0$ if it does not. The quantity $m(\tau)$ is the number of integration time
steps which make a bin.\footnote{In our simulations, the integration time
step equals $10^{-3}$ for small $N$, and $10^{-4}$ for large $N$, so that the numerical
errors are of the same order.}

Figure 5 represents the normalized average paths,
$R_\tau=(\overline{\rho}_\tau-\min \overline{\rho}_\tau)/(\max \overline{\rho}_\tau -\min \overline{\rho}_\tau)$, which start at zero in the first bin and reach 1, their maximum
value, in the central bin. The result is that the average path is temporally asymmetric, with growth
steeper than relaxation, and that, differently from the observables considered in Section 2,
it is not appreciably affected by $k^2$.

\vskip 15pt

\centerline{
\includegraphics[width=4.5cm,height=4.5cm]{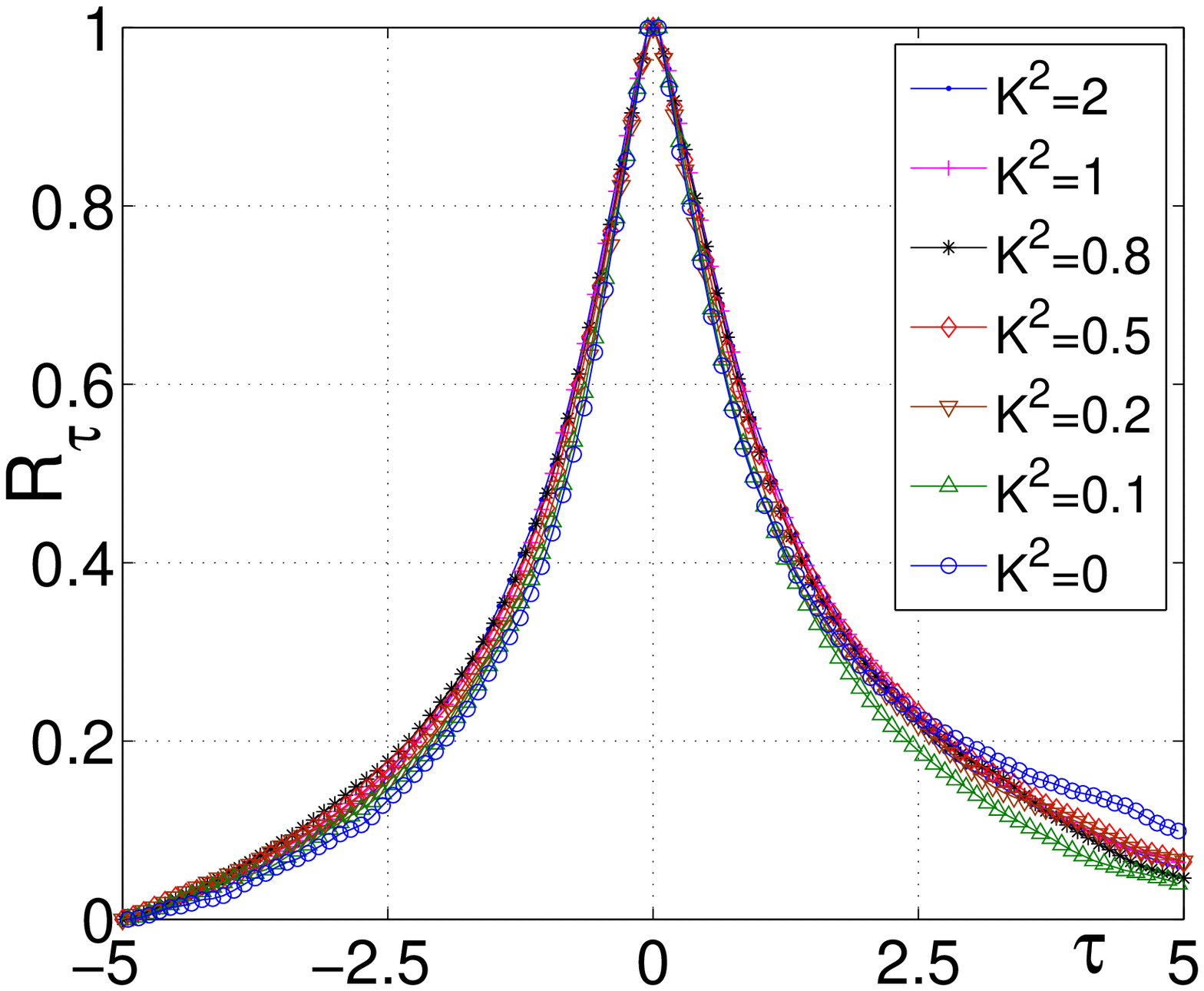}
\includegraphics[width=4.5cm,height=4.5cm]{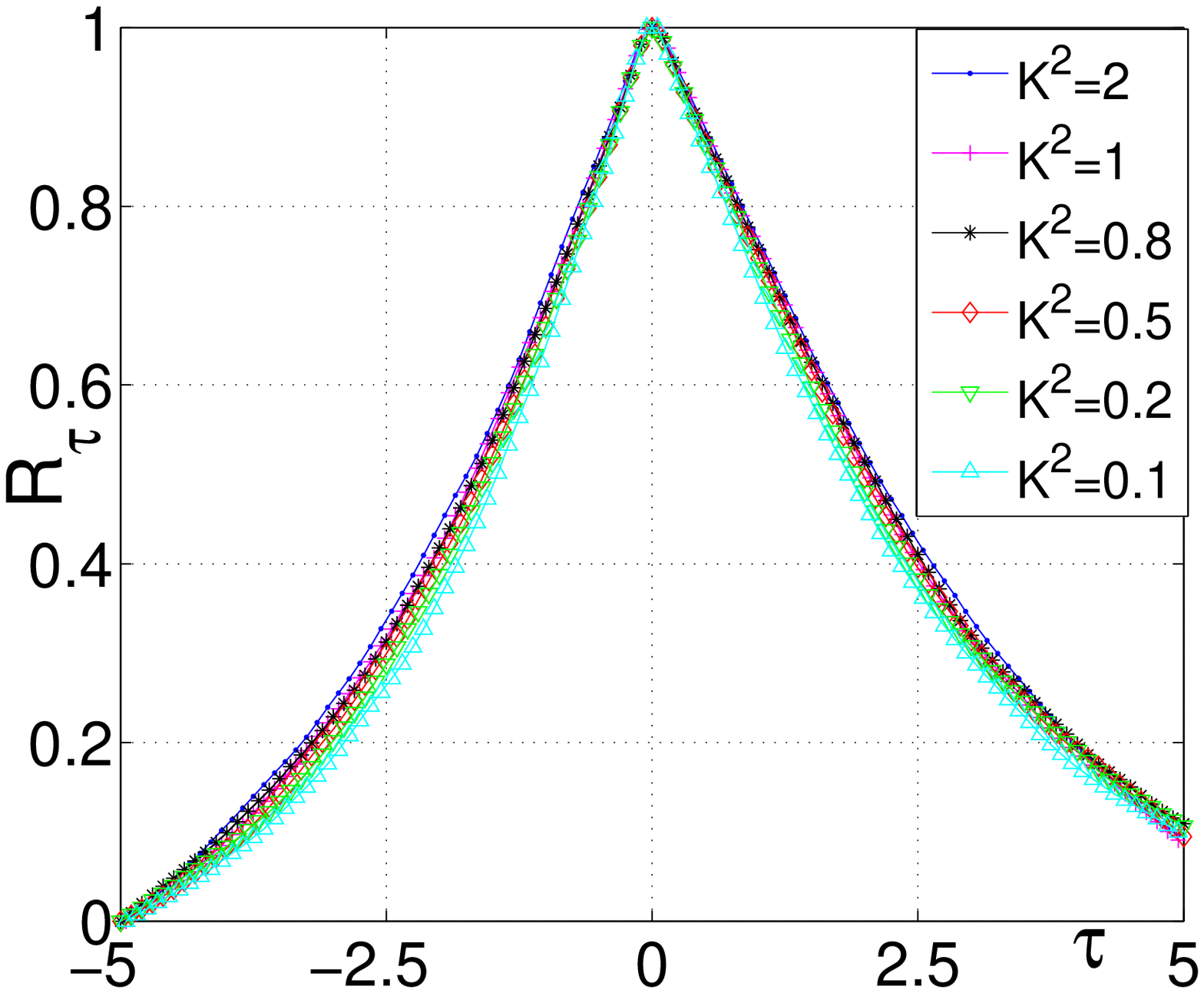}
\includegraphics[width=4.5cm,height=4.5cm]{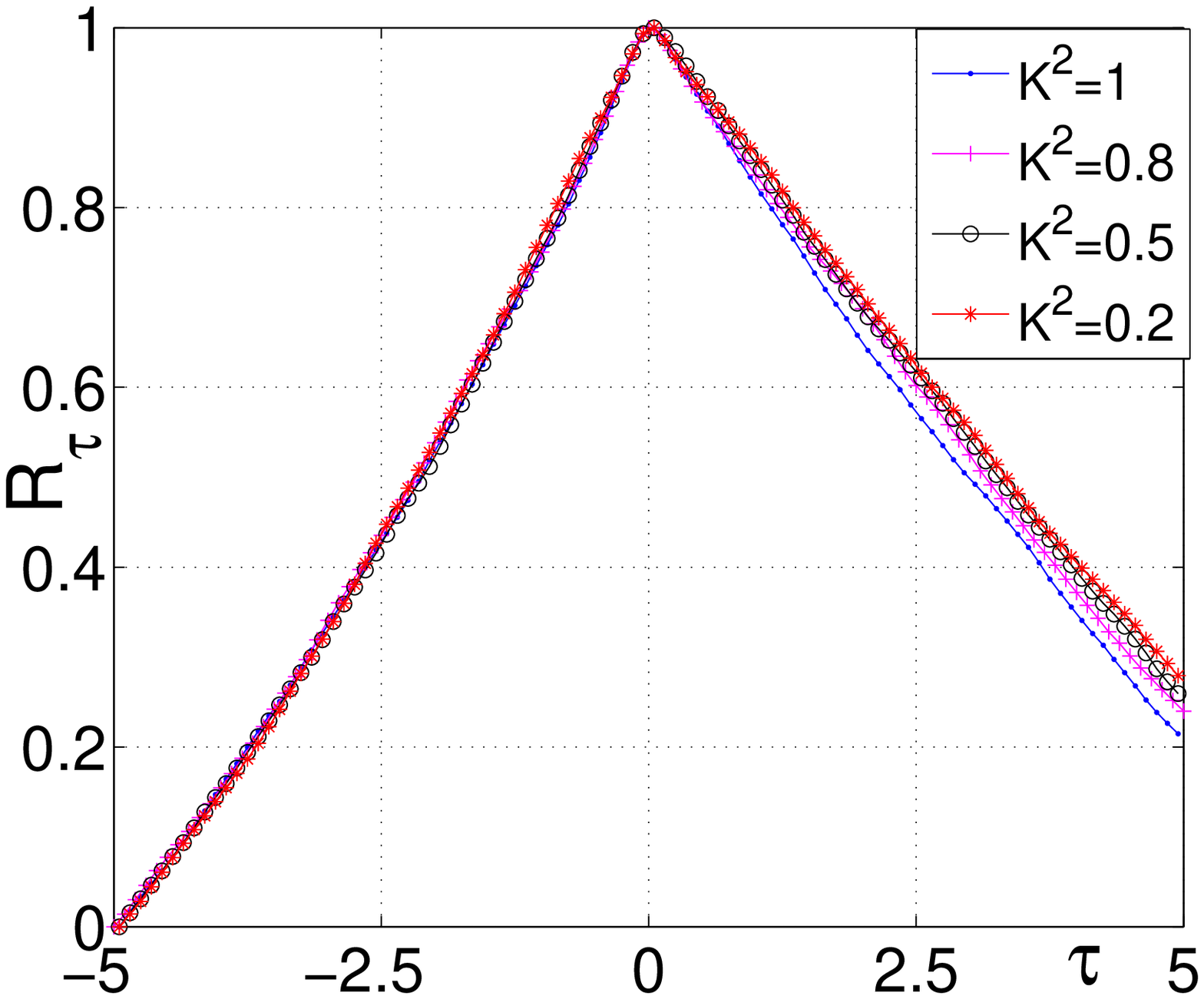}}
\vskip 3pt
{\small {\bf Figure 5.} (Color
 online) Average fluctuations of density in the
center of the chain, for different chain lengths, $N=100$ (left panel), $N=200$ (central panel), $N=400$ (right panel) and different values of the elastic constant, $k^2$.
The fluctuation paths are temporally asymmetric and not sensibly affected by the value of $k^2$.}

\vskip 15pt

This is consistent with Refs.\cite{GRV07,PSR08}, in which asymmetric paths are expected to be
typical of nonequilibrium steady states of interacting particle systems.
In Figure 5, the asymmetry seems to grow with $N$, as
in \cite{GRV07}, but a comparison of cases with different $N$ is frustrated by the unclear
dependence on $N$ of the many parameters entering the definition of the asymmetry, like $\mathcal{T}(X)$ and $t_0$. For instance, Figure 5 seems to imply that
average growth ($\zt < 0$) and relaxation ($\zt > 0$) tend to become linear as $N$
increases, but this may be due to the fact that the observation time $t_0$, which
ought to be connected with the correlations decay rate, increases with $N$.
The right panel of Figure 5 would then only illustrate a smaller fraction of the
average path than that of the left panel. As a matter of fact, Figure 6 shows that the
instantaneous asymmetry, $\delta_\tau=\left[\overline{\rho}_\tau - \overline{\rho}_{-\tau}\right]$,
almost saturates with growing $\zt$, but it saturates at larger values of $\zt$ for larger $N$.
On the other hand, it is not clear how $t_0$ should be modified with $N$; it is only
obvious that $\left[\overline{\rho}_\tau - \overline{\rho}_{-\tau}\right]$ must decrease back
to 0, for $\zt$ sufficiently long that correlations have decayed \cite{PSR08}.
Unfortunately, that time is too long to allow us to collect a statistically
relevant sample of fluctuation paths.

\vskip 15pt

\centerline{
\includegraphics[width=5.9cm]{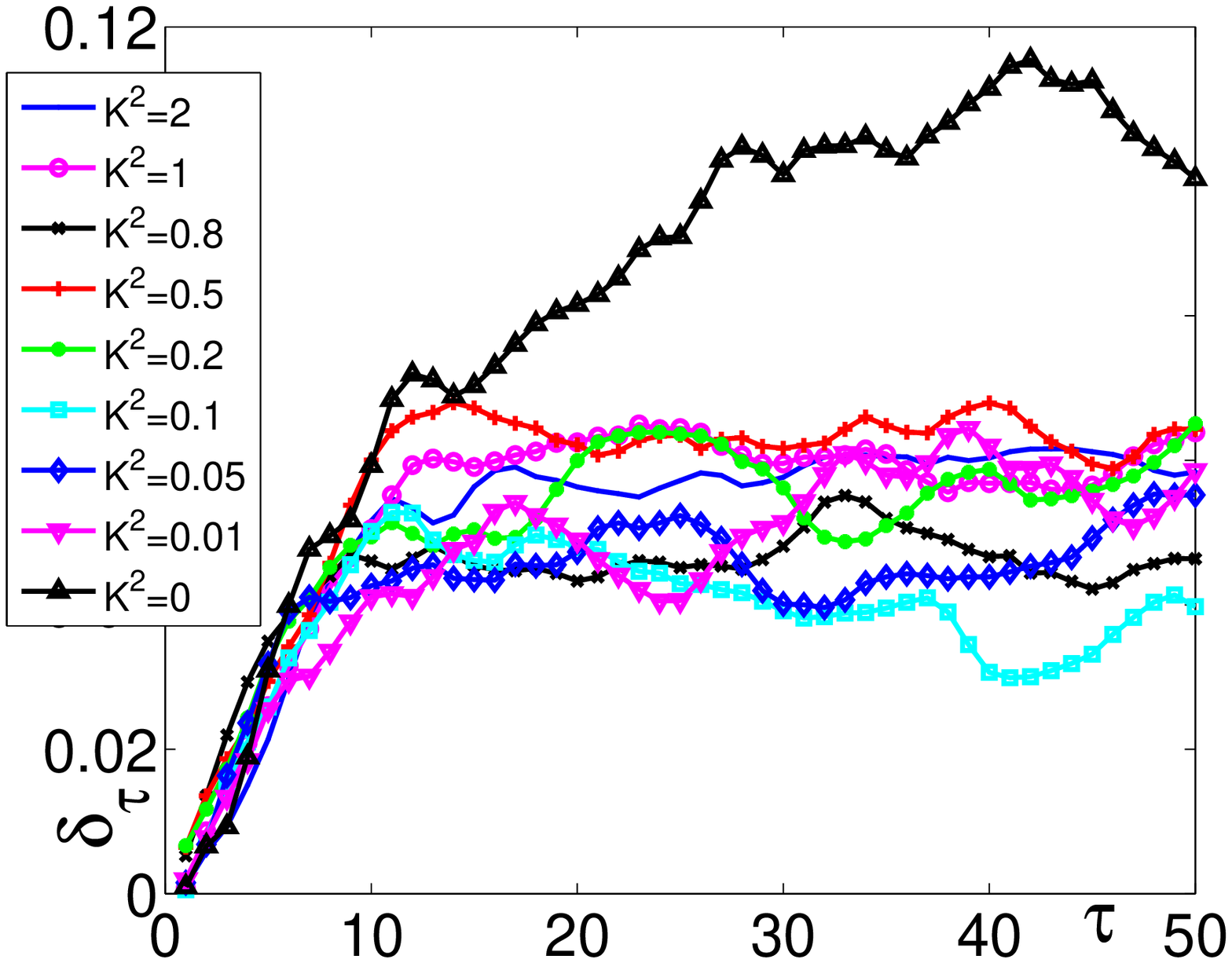} ~~
\includegraphics[width=4.6cm,height=4cm]{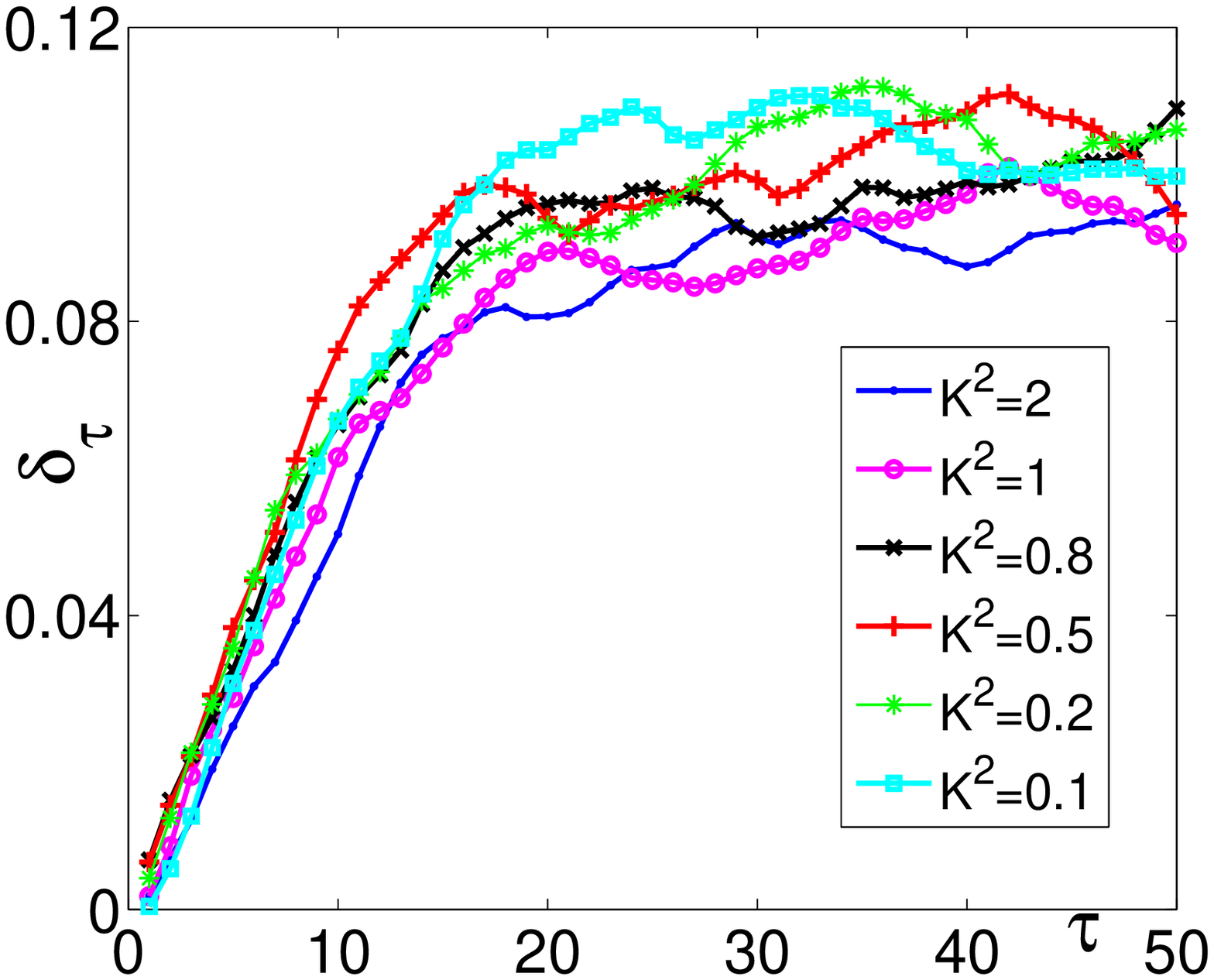} ~~
\includegraphics[width=4.5cm,height=4cm]{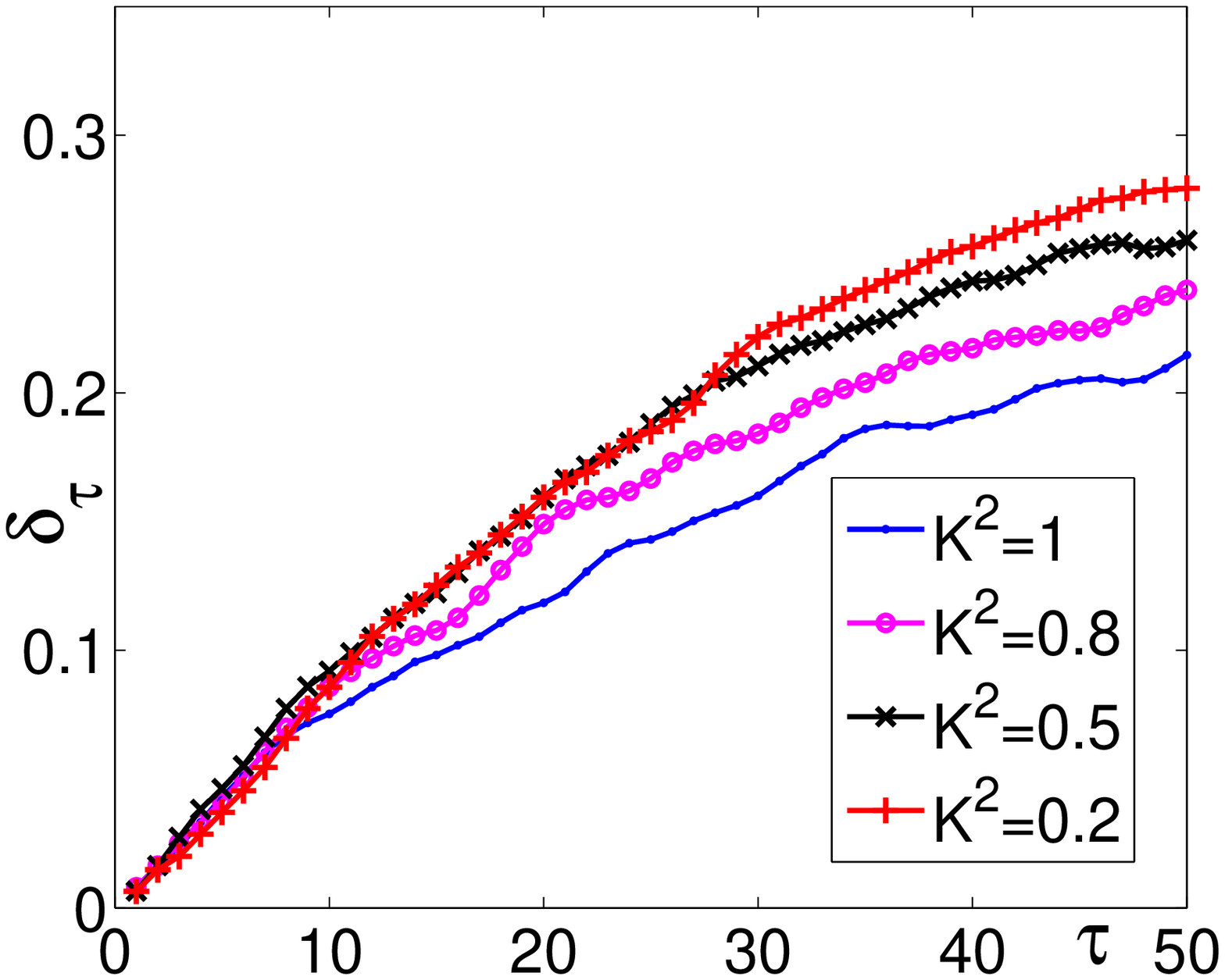}}
\vskip 3pt
{\small {\bf Figure 6.} (Color
 online) Instantaneous asymmetry of average density fluctuations,
$\delta_\tau=\left[\overline{\rho}_\tau - \overline{\rho}_{-\tau}\right]$,
for different values of the elastic constant
$k^2$, $T_\ell=320$, $T_r=20$ and $N=100$ (left panel), $N=200$ (central panel), $N=400$ (right panel). The asymmetry saturates after times which are
longer for larger $N$.}

\vskip 15pt

Let us introduce the asymmetry $\nu$, as a normalized cumulative difference between
fluctuation and relaxation trajectories,
\be
\nu = \frac{1}{\overline{\rho}_{\mbox{max}} - \overline{\rho}_{\mbox{min}}}
\left[ \sum_{t > 0} \overline{\rho}_t - \sum_{t < 0} \overline{\rho}_t \right]
\ee
The probability that this quantity be positive, $P(\nu)>0$, is reported in Table 1 for $N=100$,
where one observes that, from this point of view, the $k^2=0$ case does not differ substantially
from the $k^2>0$ cases.

\vskip 15pt

\begin{center}
\begin{tabular}{|c|c|c|c|c|c|c|c|}
  \hline
$k^2$ &    2       &    1     &    0.8    &    0.5   &    0.2    &    0.1    &      0 \\
   \hline
$P(\nu>0)$ & 0.5660    & 0.5716   &  0.5486   &  0.5710  &  0.5648   &  0.5462   &  0.5823 \\
   \hline
   \end{tabular}
\end{center}
{\small {\bf Table 1.} Probability of positive asymmetry (the fraction of
fluctuation paths with $\nu>0$), for $N=100$, $T_\ell=320$, $T_r=20$, as a
function of the elastic constant $k^2$.}

\vskip 15pt

The above figures and table suggest that nonequilibrium steady states of identical hard spheres
on a line do not behave like independent particles, although in equilibrium, when the thermostats
are removed, they do. Away from equilibrium, particles appear to develop space correlations,
which reach the bulk, eventually connecting them to the
boundaries, even in the absence of springs.
Therefore, the observed asymmetries reveal that a form of nonlocality (long range
correlations) \cite{DLS,DKS,HSpohn,ELS96} is common in nonequilibrium states of both
deterministic and stochastic models.

Another observable of interest is the heat flux. To define the heat flux in the presence
of hard core collisions, we adopt the method of planes, developed by Todd, Daivis and Evans
\cite{TDE95}, according to which the fluctuating heat flow $J_q$, through a plane transversal
to the medium, can be decomposed as the sum of two terms:
\be
J_q(x,t) = J_q^K(x,t) + J_q^U(x,t)
\ee
The term $J_q^K$ represents the kinetic part of the heat flux, which is expressed by:
\be
J_q^K(x,t) = \frac{1}{A} \sum_{i=1}^N \sum_{m_i} U_i ~ \zd(t - t_{i,m_i}) ~
\mbox{sign}[c_{x i}(t_{i,m_i})]
\ee
where $i$ denotes a particle crossing the plane at (discrete) times $t_{i,m_i}$,
$c_{x i}$ is the component along direction $x$ of the particle's velocity with
respect to the streaming velocity, $A$ is the area of the plane (if the system is 3 dimensional),
and
\be
U_i = \frac{m}{2} \left[ v_i - u(x_i) \right]^2 + \frac{1}{2} \phi_{ij}
\ee
is the internal energy concerning particle $i$. Moreover, $v_i$ is the velocity of the particle in
the laboratory frame, $u(x_i)$ is the streaming velocity at position $x_i$, and $\phi_{ij}$
is the interaction potential between particles $i$ and $j$.
The term $J_q^U$ represents the potential part of the heat flux, which is defined by:
\be
J_q^U(x,t) = - \frac{1}{4 A} \sum_{i,j=1}^N \left[ v_i - u(x_i) \right]
F_{ij} \left[ \mbox{sign}(x-x_i) - \mbox{sign}(x - x_j) \right]
\ee
where $F_{ij}$ is the force that particle $j$ exerts on particle $i$.

Figure 7 shows the average fluctuations of $J_q^U$ for the cases with
$N=100$, $T_\ell = 320$, $T_r=20$, and for different
values of $k^2$. Analogously to the density average path, the average $J_q^U$ fluctuation
rises faster than it relaxes, and is not appreciably affected by the value of $k^2$.
Results similar to those reported have been obtained for all different parameters choices,
even with $k^2=0$, when the bulk behaviour should have closely resembled that of
non-interacting particles.
This is consistent with the discussion of the single file behaviour of \cite{JBR08}, according to
which the order of particles introduces persistent correlations, even in the presence of positive Lyapunov exponents. The results appear to be robust and independent of the time reversal
parity of the observable at hand, since local density and heat flux have opposite parity.

The results of this section definitely prove that asymmetric fluctuation paths are typical
of deterministic, time reversal invariant nonequilibrium models.

\vskip 15pt

\centerline{\includegraphics[width=6cm]{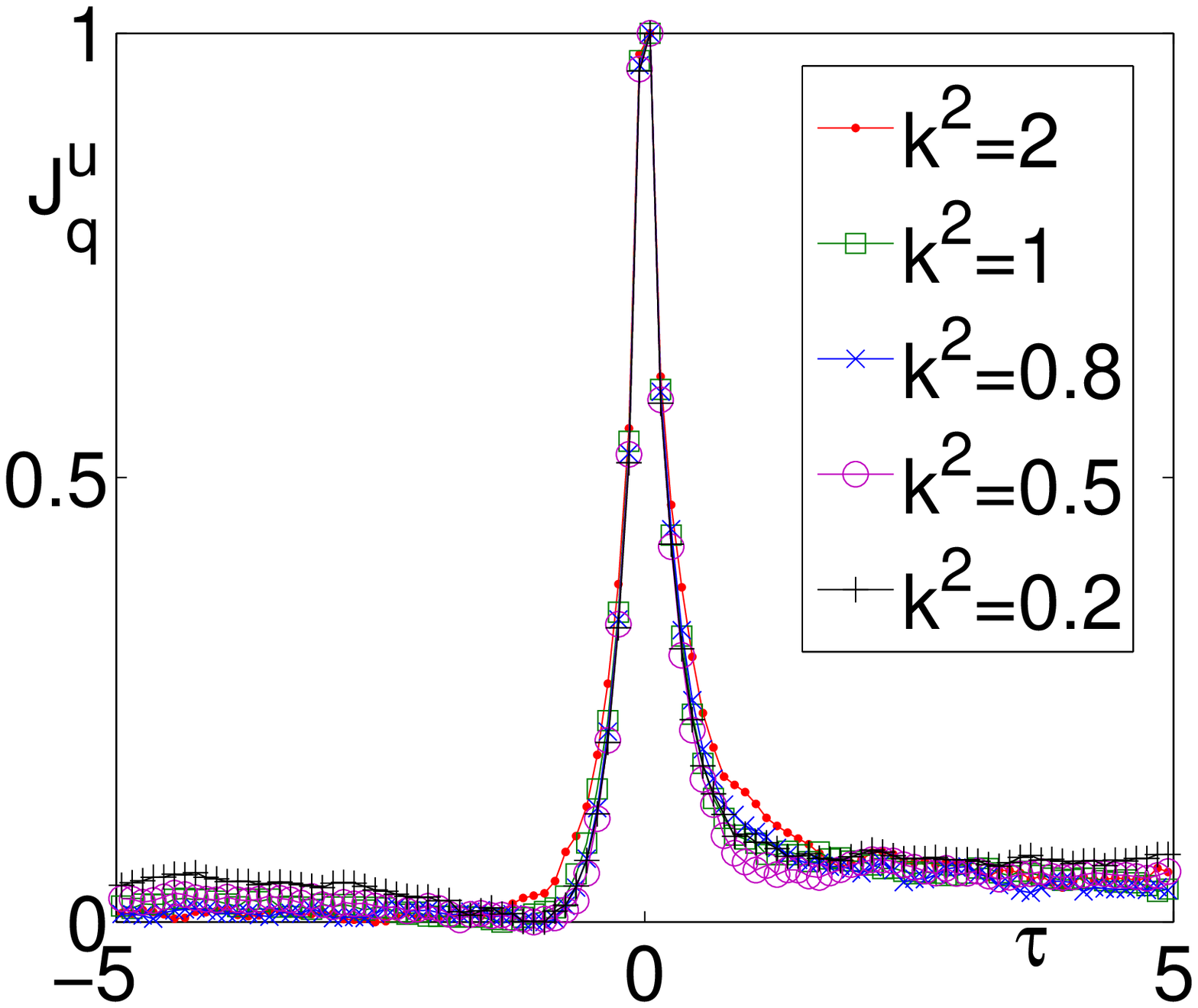}}
\vskip 3pt
{\small {\bf Figure 7.} (Color
 online) Average fluctuations of
$J_q^U$, for various values of $k^2$, with $N=100$, $T_\ell = 320$, $T_r=20$ and $\theta_\ell=\theta_r=1$.
The fluctuations have been rescaled, so that they all vanish at time $-\zt$, and take the
peak value 1 at time 0.}

\newpage

\section{Harmonic chain in a nonequilibrium steady state}
\label{Harmonic}
In this section we consider the model studied by Rieder, Lebowitz and Lieb
\cite{RLL}, with the stochastic boundary thermostats replaced by Nos\'e-Hoover
thermostats, which make the dynamics time reversal invariant. The oscillators are
thus coupled by the potential energy $V$ of Eq.(\ref{potential}), with $\zb=0$. In equilibrium,
this system is equivalent to a set of independent particles, the normal modes.

If applicable, the local version of the virial theorem now implies \cite{LLP1}:
\begin{equation}
\langle p_i^2\rangle=- \langle (F_i-F_{i+1})q_i\rangle,\quad i=2,\ldots, N-1 ~.
\label{LocVir}
\end{equation}
Clearly, in the absence of thermostats, the local virial theorem does apply to each
independent mode of oscillation, but in general, this has to be checked.
As in the case of stochastic baths considered by \cite{RLL}, we find that
Eq.(\ref{LocVir}) holds and that the kinetic temperature profile is characterized by jumps
at the boundaries, and by approximately flat profiles in the bulk. As in \cite{RLL}, harmonic
chains do not sustain a kinetic temperature gradient but, differently from the
case of \cite{RLL}, the approximate equipartition of energy in the bulk does not correspond
to the mean of the left and right temperatures: it stays closer to the higher temperature,
cf.\ Fig.8.

\vskip 15pt

\centerline{
\includegraphics[width=8.5cm,height=6cm]{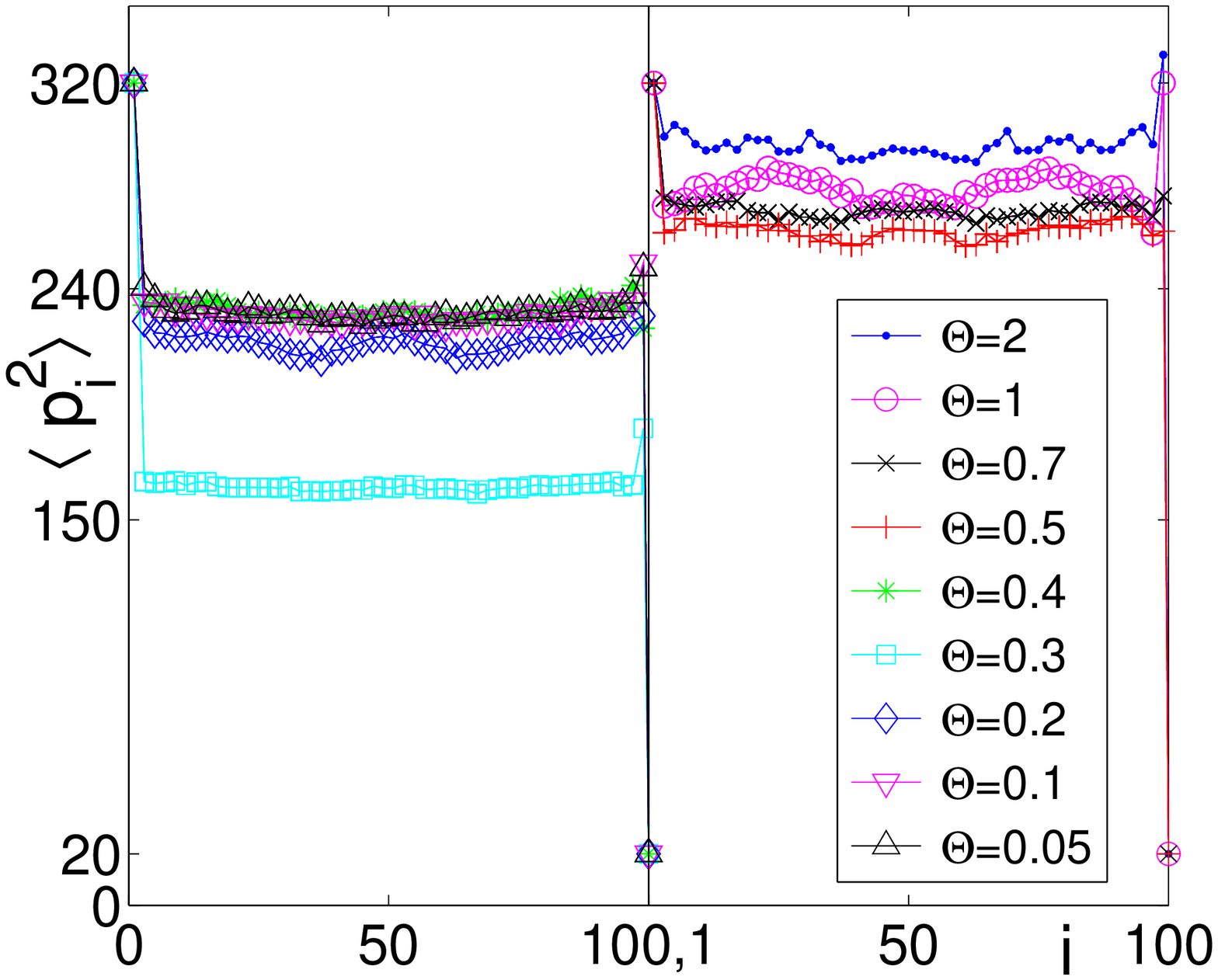} ~
\includegraphics[width=7.5cm,height=6cm]{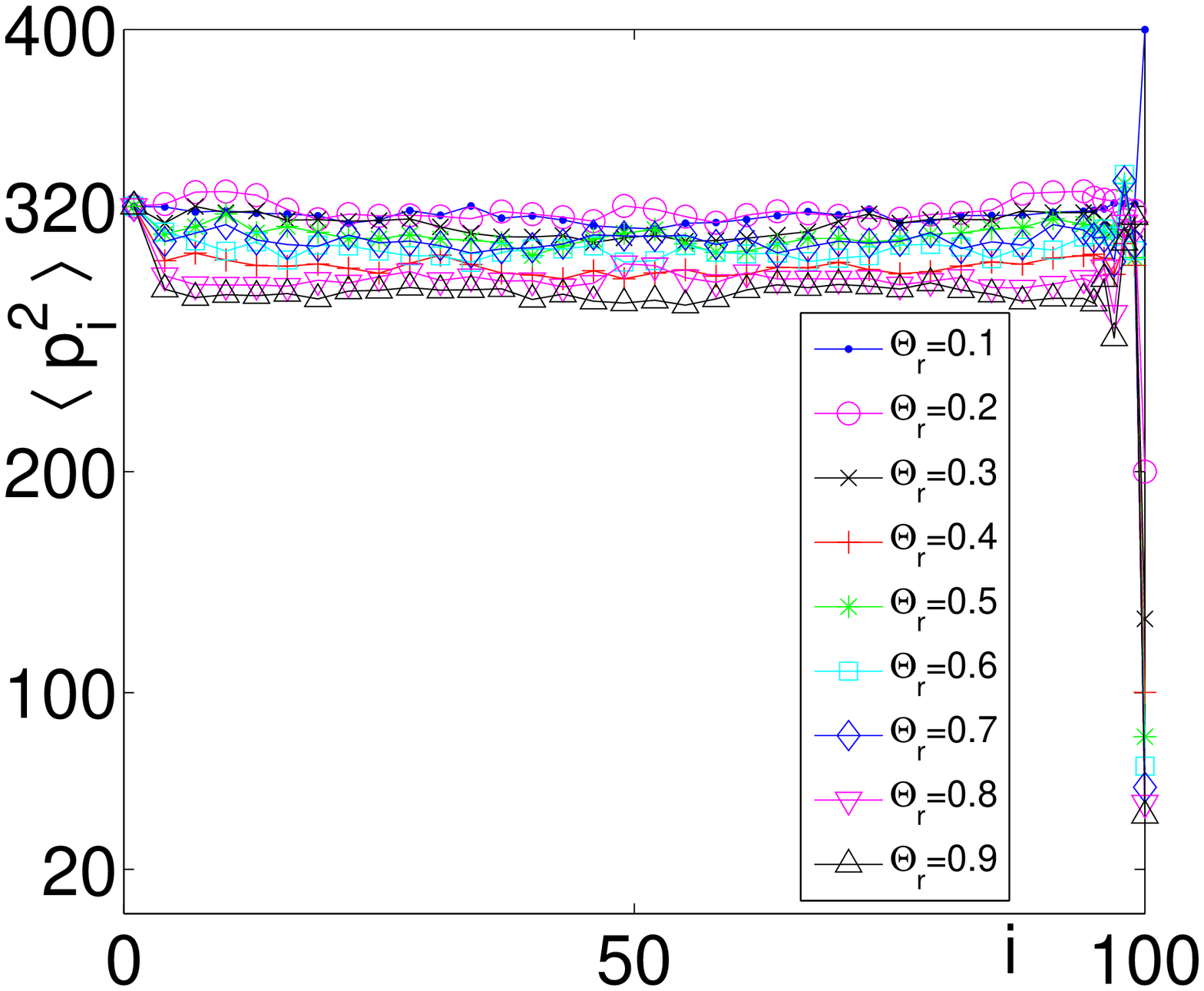}} \noi
{\small {\bf Figure 8.} (Color
 online) Temperature profile as a function of the thermostat characteristic
time $\theta$. The shorter $\theta$, the stronger the coupling between particle and thermostat.
Results at short $\theta=\theta_\ell=\theta_r$ are reported in the left half of the left panel, while results at large
$\theta$ are reported in the right half. After some oscillation, decreasing $\theta$ makes the profile
settle around the value 240, which is much closer to $T_\ell=320$ than to $T_r=20$.  Note the apparent discontinuity of the profile between $\theta=0.5$ and $\theta=0.3$, which leads to the flat bulk
profile at $(T_\ell+T_r)/2$. The right panel portrays the profiles with $\theta_\ell = 1$
and decreasing $\theta_r$: the system equilibrates with the hot thermostat
independently of $\theta_r$. In both panels $N=100$.}

\vskip 15pt

Some dependence of the bulk behaviour on the characteristics of the thermostats is not
surprising: the choice of the parameters determines the efficiency of the energy exchange
between thermostats and thermostatted system. This,
in turn, may affect the behaviour of a non-macroscopic system since, by definition, this
does not enjoy any local thermodynamic equilibrium.
In our case, the relevant parameters consist of the target temperatures $T_\ell$ and $T_r$,
of the thermostat characteristic times $\theta_\ell$ and $\theta_r$, and of the elastic
constant $k^2$. We have studied the behaviour of the profile in the
$\theta=\theta_\ell=\theta_r \to 0$ limit, the limit of strong coupling between
thermostats and thermostatted particles, as well as in weak coupling cases.
Figure 8 (left panel), shows that the
kinetic temperature profile settles closer to $T_\ell$ than to $T_r$, at all values
of $\theta$, except at $\theta=0.3$. The situation does not change if $\theta_r$ is
shorter than $\theta_\ell$, so that the cold thermostat is more strongly coupled to the system
than the hot thermostat, as illustrated by the right panel of Fig.8.
For $\theta=0.3$, the deterministic time
reversal invariant Nos\'e-Hoover thermostats behave most similarly to the irreversible
stochastic ones of \cite{RLL}, as far as the kinetic temperature profile is concerned.
Note, in particular, the flatness of the $\theta=0.3$ profile, and the temperature
overshoot on the cold side, predicted by the theory of \cite{RLL}. The equivalence with
the results for stochastic thermostats of \cite{RLL} is not complete, however,
because the temperature does not overshoot in the hot side.

\vskip 15pt

\centerline{
\includegraphics[width=7.5cm,height=6cm]{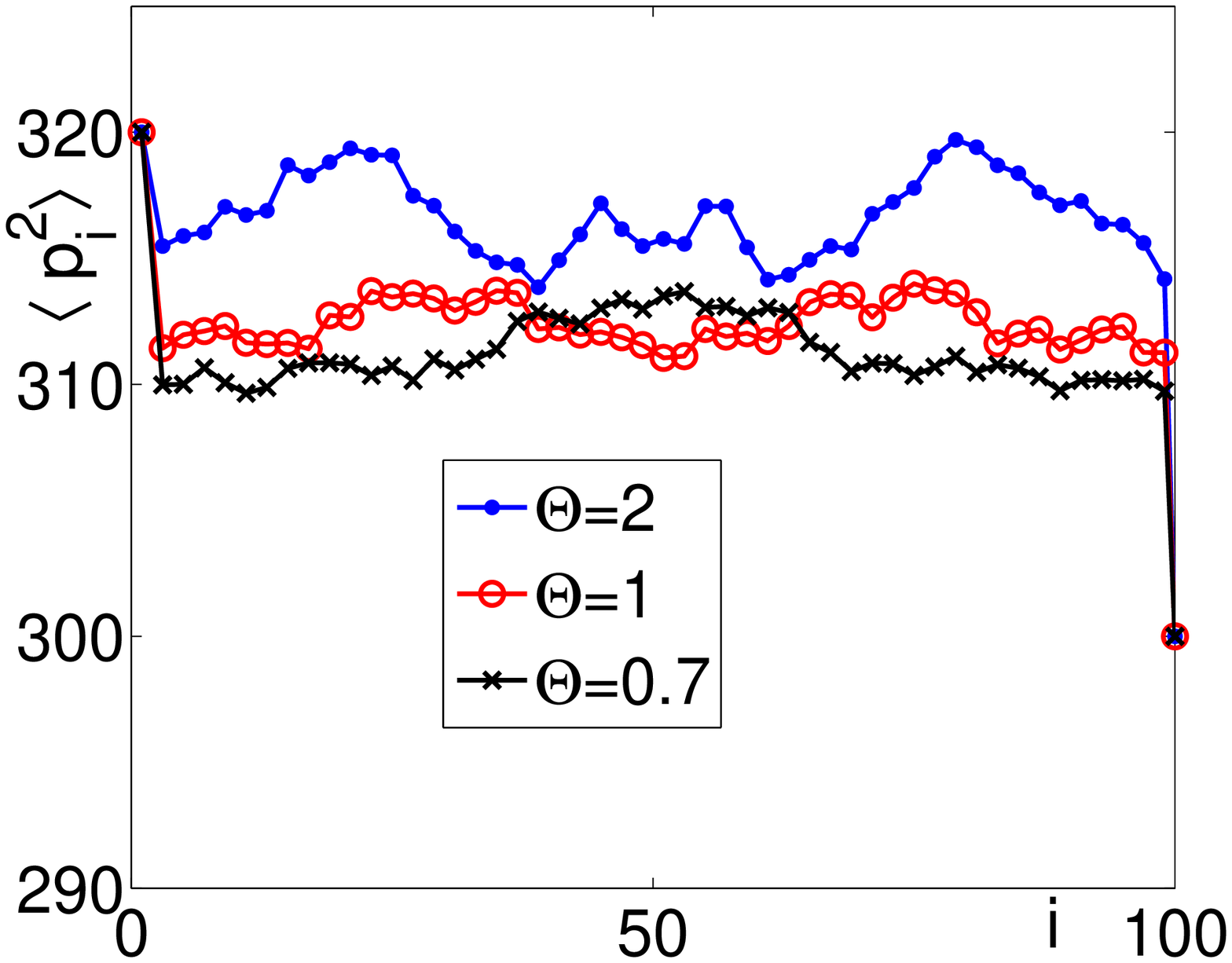}
\includegraphics[width=8.5cm,height=6cm]{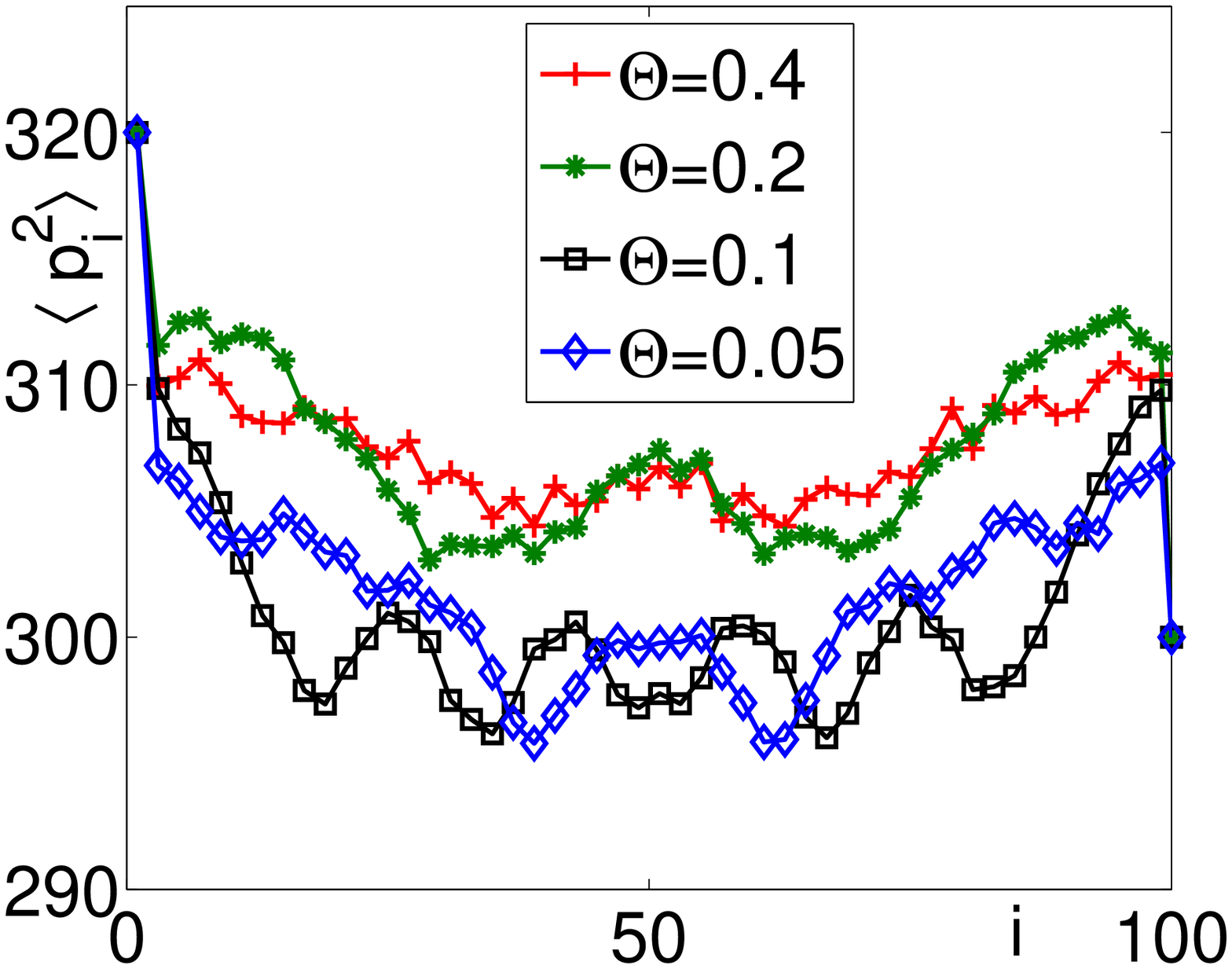} } \noi
{\small {\bf Figure 9.} (Color
 online)
Temperature profile as a function of $\theta=\theta_\ell=\theta_r$, for $N=100$,
$T_\ell=320$, $T_r=300$. For high $\theta$. i.e.\ weak coupling, it settles
within $[T_r,T_\ell]$, while for low $\theta$, i.e.\ strong coupling, it reaches even
lower than $T_r$. The profile depends quite irregularly on $\theta$. For $\theta=0.7$
and $\theta=1$ the bulk temperature settles close to $(T_\ell+T_r)/2$.}

\vskip 15pt
This is consistent with the observations of Ref.\cite{MMR08}, on a system of rotating
disks and pointlike particles, thermostatted by Nos\'e-Hoover mechanisms acting on
the boundary disks. There, the hotter side reaches local equilibrium more
easily than the colder one, even if not necessarily at the target temperature of the
thermostat. The thermostat closer to an equilibrium state (as it should be)
seems, then, more efficient in driving the bulk of the
chain than the other thermostat. In Ref.\cite{MMR08}, it was further observed
that small gradients lead to standard behaviour of the cold thermostat. Differently,
in our case, the temperature profiles remain quite irregular even under relatively small
gradients, and only for a limited set of $\theta$'s do they lie close to $(T_\ell+T_r)/2$.
The profiles worsen in the small $\theta$ limit, i.e.\ for strong couplings, when
they even brim over the interval $[T_r,T_\ell]$, cf.\ Fig.9.
The strong dependence
on the microscopic parameters is confirmed by the departure at large $\theta$ of the
bulk profiles from the mean of the boundary values, cf.\ the case $\theta=2$ in the
left panel of Fig.9.

%\vskip 15pt
%
%\centerline{
%\includegraphics[width=5cm]{distribp1N100Tl320Tr20Theta03solomolle_bis.eps}
%\includegraphics[width=5cm]{distribp50N100Tl320Tr20Theta03solomolle_bis.eps}
%\includegraphics[width=5cm]{distribp100N100Tl320Tr20Theta03solomolle_bis.eps}}
%\vskip 3pt
%{\small {\bf Figure 10.} Histograms of velocities for particles 1, 50 and 100, in a purely
%harmonic, themrostatted chain with $N=100$, $T_\ell=320$, $T_r=20$, and $\theta=0.3$.}
%
%\vskip 15pt

To further assess the relation between our chains and those of \cite{RLL} we have produced
the histograms of $p_i$, and have verified that they are not Gaussian, although to
different degrees for the different $i$'s. Indeed, the distributions of $p_1$
and $p_{N/2}$ differ much less from a normal distribution than that of $p_N$ ($N$ is the particle
interacting with the cold bath at temperature $T_\ell)$, as indicated
by normal distribution plots not reported here. Also, local Gaussian distributions
are better and better approximated at both ends of the chain, as in
\cite{MMR08}, if the temperature gradient is reduced. Increasing $k^2$, without
changing the thermostats properties, is another route towards approximately
Gaussians local distributions.
%This is illustrated by a comparison of Figure 10 with Figure 11,
%where the normal probability plots for particles $i=1$ and $i=100$ are also given (a straight line %means Gaussian distribution).
This confirms that closer to equilibrium it is easier to obtain the equivalence between deterministic and stochastic thermostats, as generally expected \cite{JR10}, although
the temperature profiles demonstrate that complete equivalence is not to be expected.

%\vskip 15pt
%
%\centerline{
%\includegraphics[width=5cm]{distpN100K1Tr320Tl300solomolle.eps}
%\includegraphics[width=5cm]{probpltp1N100K1Tr320Tl300solomolle.eps}
%\includegraphics[width=5cm]{probpltp100N100K1Tr320Tl300solomolle.eps}}
%\vskip -1pt
%{\small {\bf Figure 11.} Histograms of velocities for particles 1, 50 and 100 (left) in a purely
%harmonic, thermostatted chain with $N=100$, and small temperature
%gradient: $T_\ell=320$, $T_r=300$.
%The central and right panels depict the log-normal probability plots for the
%histograms of the first
%and last particle.}
%\vskip 15pt

The peculiar behaviour of the purely harmonic nonequilibrium chains is evidenced also by quantities
such as $\langle x_{i+1} - x_i \rangle$ and $\langle q_i \rangle$, whose profiles are much more
irregular than those of Figure 3, cf.\
Figure 10. Here, differently from Figure 3, $\langle q_i \rangle$ almost vanishes, consistently
with the approximately equilibrium state of the bulk of the chain.\footnote{Note that this state
only looks close to equilibrium: for instance, position-velocity correlations differ
significantly from their equilibrium counterparts \cite{LLP08}.}

\vskip 15pt

\centerline{\includegraphics[width=7cm]{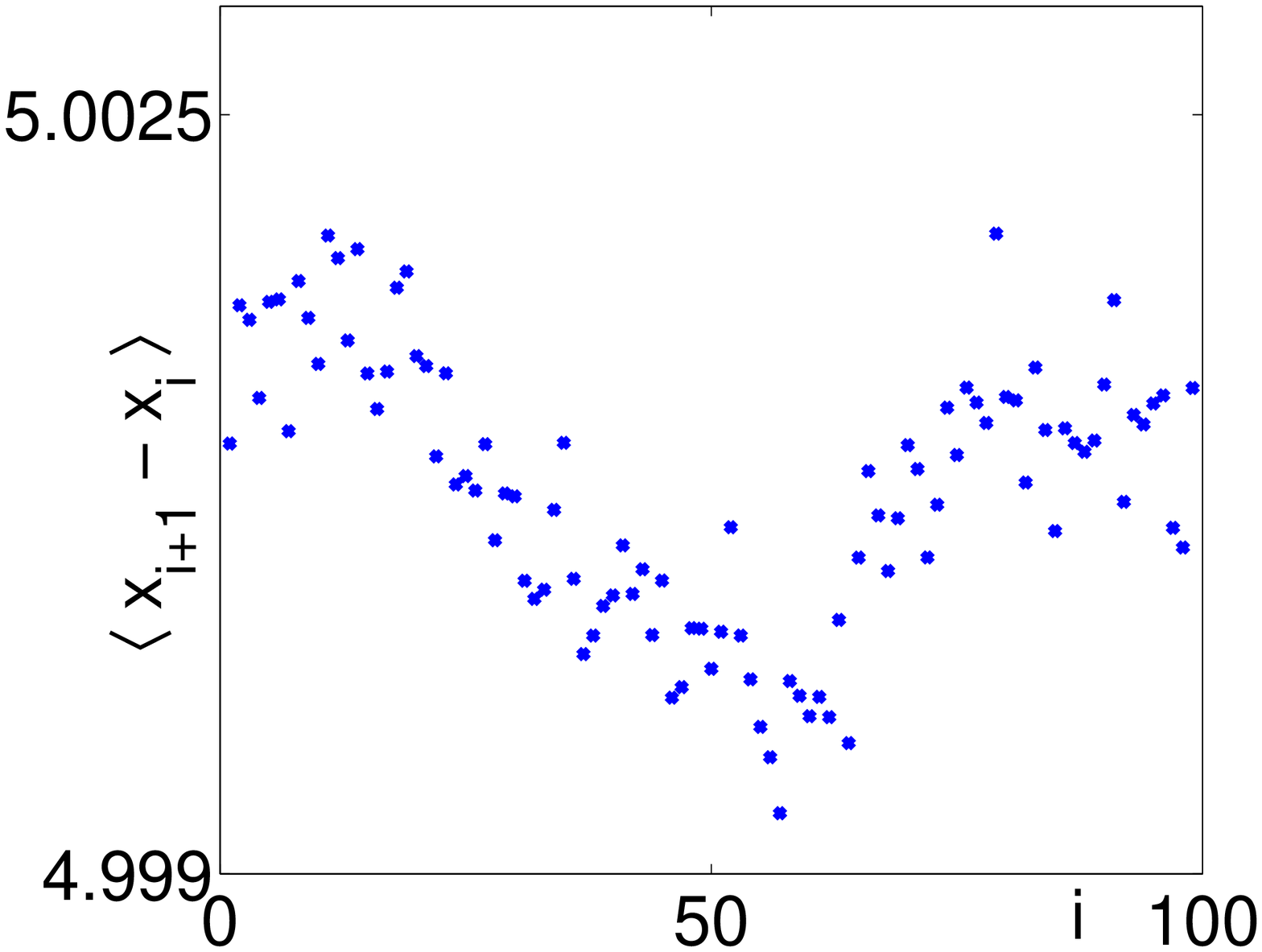} ~~
\includegraphics[width=7cm]{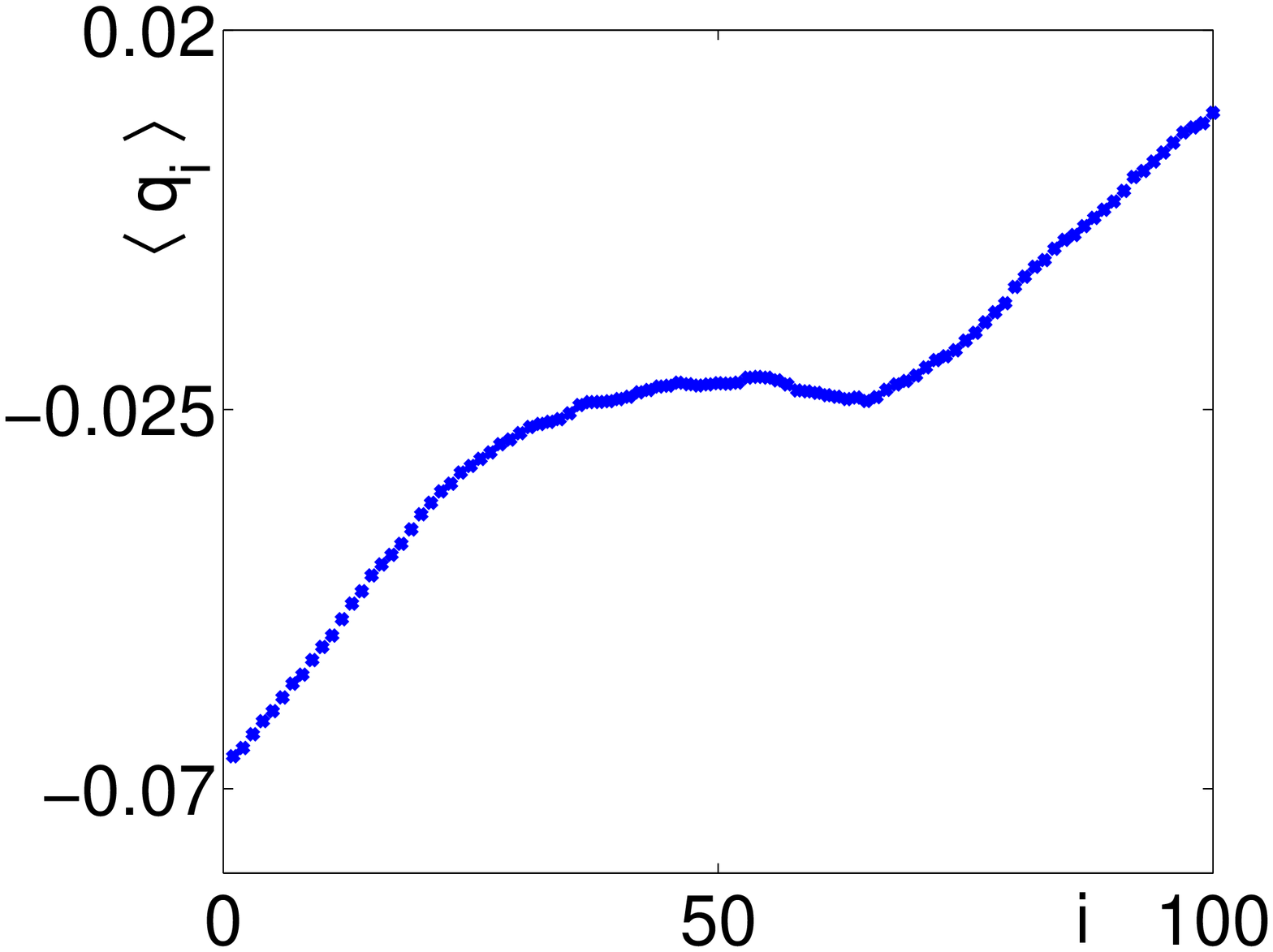}}
\vskip -4pt \noi
{\small {\bf Figure 10.} (Color
 online) Average coordinates difference between nearest neighbours,
$\langle x_{i+1} - x_i \rangle$, for $k^2=1$ and equilibrium distance $a=5$ (left panel).
Average displacement $\langle q_i \rangle$ from the equilibrium position, $ai$ (right panel).
In both panels $N=100$, $T_\ell=320$, $T_r=20$, $k^2=1$.}

\newpage

\section{Lyapunov exponents: the harmonic and the hard core cases}
In this section we consider the Lyapunov exponents of the two limiting cases of purely
harmonic and purely hard core interactions.
These exponents have been computed in tangent space, with the usual algorithm
devised by Benettin, Galgani, Giorgilli and Strelcyn \cite{BGGS}. The phase space
contraction rate $\chi$, i.e.\ the dynamical dissipation, can be separately computed
as the negative of the average of the divergence of the vector field, neglecting the
instantaneous elastic collisions,
which do not contribute to the variations of the phase space volumes \cite{BGG}.
In our case,
$\langle \chi \rangle = (\langle\xi_\ell\rangle+\langle\xi_r\rangle)$ and
$\langle \chi \rangle \ge 0$, where equality characterizes equilibrium states,
while $\langle \chi \rangle > 0$ characterizes nonequilibrium states.

The motion of purely harmonic chains, in the absence of thermostats, is hamiltonian
and fully integrable, hence all Lyapunov exponents vanish. Table 2\
shows the values of the largest Lyapunov exponent, $\zl_1$, for several different
kinetic temperatures at the left end of a chain of 100 particles, while the right end is
subjected to a thermostat with $T_r=20$.

\vskip 15pt

\begin{center}
\begin{tabular}{|c|c|c|}
  \hline
  $T_\ell$ & $\lambda_1$ & $\langle\chi\rangle$ \\
  \hline
   420 &    $0.00771 \pm 0.5\times 10^{-4}$ &  2.7988 \\
   \hline
   370  &    0.00769 $\pm 0.6\times 10^{-4}$ & 2.5737   \\
   \hline
   320&  0.00779  $\pm 0.7\times 10^{-4}$ &  2.3382 \\
  \hline
   220 &  0.00792 $\pm 0.5\times 10^{-4}$  &  1.7644 \\
     \hline
   120 &   0.00928$\pm 0.1\times 10^{-3}$  & 1.0726 \\
     \hline
    80 & 0.01101 $\pm 0.1 \times 10^{-3}$ & 0.7194  \\
      \hline
    50 &   0.01338 $\pm 0.2 \times 10^{-3}$& 0.3775 \\
      \hline
    20 &    0.01328 $\pm 0.1\times 10^{-3}$  & $6.54\times 10^{-5}$  \\
   \hline
   \end{tabular}
\end{center}
{\small {\bf Table 2.} Largest Lyapunov exponent, $\zl_1$, and average phase space contraction
rate, $\langle\chi\rangle$, as functions of $T_\ell$, for purely harmonic chains with
$N=100$, $k^2=1$ and $T_r=20$.
Standard deviations of the computed values are reported, with the number in brackets representing
the power of 10 which multiplies them.}

\vskip 15pt

It is interesting to note that even at equilibrium, i.e.\ for $T_\ell = T_r$,
the presence of the thermostats makes chaotic the dynamics. Indeed, $\zl_1$ is higher
at lower temperature differences, since higher dissipations imply smaller attractors
and more orderly states. It also appears that $\zl_1$ saturates as a function of the
temperature difference, while the dissipation keeps increasing significantly.
As a function of the relaxation times of the thermostats, the largest Lyapunov exponent
decreases monotonically. For instance, the largest exponent for $N=100$ and $T_\ell=320$
decreases from the value $\zl_1 = 0.00779$ reported in Table 2 for $\theta=1$ to
$\zl_1 = 0.000444$ for $\theta=10$. This is a consequence of the fact that the motion
of single oscillators is integrable and that large $\theta$ practically decouples the
thermostats from the system of interest.
Table 3 reports the computed values of $\zl_1$ and $\chi$, at fixed temperature
difference and varying chain length. The Lyapunov exponent decreases with $N$, hence with
the temperature gradient, while the dissipation $\langle \chi \rangle$ is
practically constant.

\vskip 15pt

\begin{center}
\begin{tabular}{|c|c|c|}
  \hline
  $N$ & $\lambda_1$ & $\langle \chi \rangle$ \\
  \hline
  50 & 0.0120 $\pm 0.8\times 10^{-4}$ &  2.3013\\
    \hline
  100& 0.00779 $\pm 0.7 \times 10^{-4}$ & 2.3382 \\
  \hline
  150 & 0.0062 $\pm 0.8\times 10^{-4}$ & 2.3286 \\
  \hline
  200 & 0.0048 $\pm 0.1\times 10^{-3}$ &  2.3296 \\
  \hline\hline
\end{tabular}
\end{center}
{\small {\bf Table 3.} Largest Lyapunov exponent and dissipation for the purely harmonic case,
with $k^2=1$, and fixed temperature difference: $T_r=20, T_\ell=320$.
The global gradient decreases with $N$, like $\zl_1$, while $\langle \chi \rangle$ is
practically constant.}

\vskip 15pt

The picture is completed by the behaviour with $N$ at fixed temperature gradient,
cf.\ Table 4.\footnote{As observed in  e.g.\ \cite{AK03}, the local kinetic temperature gradient
inside the chain is not a useful variable, in this case, because it fluctuates.}
The data indicate that two competing effects contribute to the values of the largest
Lyapunov exponent:
\begin{itemize}
\item[{\bf a)}]
decreasing the temperature gradient, the system approaches
the equilibrium state, which enjoys the largest value of $\zl_1$ afforded by a chain
of given length;
\item[{\bf b)}]
increasing $N$ reduces the impact of the thermostats on the bulk of the chain,
which then better approximates an isolated, fully integrable system, with vanishing
Lyapunov exponents.
\end{itemize}
As $\zl_1$ does not grow indefinitely with $(T_\ell - T_r)$, increasing $N$
leads to a decrease of $\zl_1$. The dissipation $\langle \chi \rangle$, on the other hand,
is more directly related to the value of $(T_\ell - T_r)$, and grows with it.

\vskip 15pt

\begin{center}
\begin{tabular}{|c|c|c|c|c|c|c|}
  \hline
  $N$ & $\lambda_1$ & $\langle \chi \rangle $\\
  \hline
  50  & 0.0131 $\pm 0.68\times 10^{-4}$ &  1.4384\\
  \hline
  100 & 0.0078 $\pm 0.73\times 10^{-4}$ & 2.3382 \\
  \hline
  150 & 0.0057 $\pm 0.96\times 10^{-4}$ &  3.0205 \\
  \hline
  200 & 0.0050 $\pm 0.67\times 10^{-4}$ &   3.6501 \\
  \hline\hline
   \end{tabular}
\end{center}
{\small {\bf Table 4.} Largest Lyapunov exponent and dissipation for the purely harmonic case,
with $k^2=1$, $T_r=20$ and fixed global temperature gradient
$(T_\ell - T_r)/N = 3$.}

\vskip 15pt

An interesting feature of the purely harmonic case, with Nos\'e-Hoover thermostats at
different temperatures, is that non-chaotic dynamics can at least be realized in short chains.
For instance, in the case with $N=3$, $T_\ell=80$, $k^2=1$ and $T_r=20$, all Lyapunov exponents
are negative, except for one vanishing exponent, and the motion is periodic.
As a matter of fact, even some cases with larger $N$ might share these features, but
their largest exponents appear to be only marginally positive, hence hard to distinguish
from vanishing values. Differently, the case with $N=4$ has clearly
positive exponents.

Consider now the case in which particles interact via hard core elastic collisions only, which,
in the absence of thermostats, would be the other limiting, fully integrable case.
The largest exponents are commonly found to be positive and behave as in
the purely harmonic case, cf. Table 5.

\begin{center}
\begin{tabular}{|c|c|c|}
  \hline
  $T_\ell$ & $\zl_1$ & $\langle \chi \rangle$ \\
  \hline
   \hline
   320 &0.01641$\pm 0.16\times 10^{-2}$ &  8.41388  \\
  \hline
   220 &0.01931$\pm 0.18\times 10^{-2}$ &  4.93916  \\
     \hline
   120 &0.01978$\pm 0.10\times 10^{-3}$  &  2.05788 \\
     \hline
    80 & 0.02441 $\pm 0.14 \times 10^{-3}$ & 1.02030 \\
      \hline
    50 &   0.02983 $\pm 0.22 \times 10^{-3}$ & 0.33846 \\
      \hline
    20 &0.03157  $\pm 0.16\times 10^{-3}$  &  0.04161 \\
   \hline
   \end{tabular}
\end{center}
{\small {\bf Table 5.} Largest Lyapunov exponent and dynamical dissipation for hard particles with
$k^2=0$, $N=100$ and $T_r=20$. The behaviours of both
$\zl_1$ and $\langle \chi \rangle$ do not differ substantially from those of purely harmonic cases.}

\vskip 15pt
The comparison between hard core and purely harmonic cases illustrates
one interesting fact, consistent with the observations of Ref.\cite{JR06}. Both kinds of systems
have positive Lyapunov exponents, but the purely harmonic case shows a quite peculiar bulk
equipartition of energy, while the chains of purely hard core particles verify more standard nonequilibrium conditions. The randomness entailed by
hard collisions, which occur at practically random times and positions,
as assumed in \cite{LMMP09}, breaks correlations among particles more efficiently than generic
chaotic mechanisms do and favours standard behaviour, as argued also in \cite{JR06}.
However, the other peculiarities noted in the previous sections still do not allow us to speak
of genuine local thermodynamic equilibrium for chains of hard particles. That chaos does
not suffice for standard behaviour, in FPU systems, has been observed in the
past \cite{LLP1,LLP97}.

\newpage

\section{Concluding remarks}
We have analyzed deterministic chains of oscillators with harmonic forces and elastic
collisions, to asses the equivalence of deterministic time reversal invariant and
stochastic models of
thermostats. Close to equilibrium, it is easier to obtain equivalent behaviours
than far from equilibrium, although the peculiarities of 1-d dynamics prevent
complete equivalence, even in the large $N$ limit. Only from a qualitative
standpoint, various properties are common to the different models, like the
form of the temperature profiles, the temporal asymmetries of flucutations
and the nonlocality of nonequilibrium steady states \cite{BDGJL2007,bsgj01,BDSJLcurrent,DLS,HSpohn}. Similarly, the strong dependence of
the bulk behaviour on microscopic details of the dynamics, such as the boundary conditions,
which is enhanced by the growth of the dissipation, reveals qualitative similarities
between the different kinds of thermostatted evolutions.
This qualitative similarity is robust against changes of the interaction potentials. For
instance, the overall behaviour of systems made of purely harmonic identical oscillators
differs from that implied by hard core and other non-linear forces, but this happens for
both stochastic and deterministic thermostats.

Therefore, from relatively coarse, qualitative, viewpoints may we state that the different
models of thermostat are equivalent.
This weak equivalence suffices
to conclude that anomalous behaviours are in fact typical of physics in (quasi) 1-dimension, but
does not suffice to predict the kind of anomalies that one should expect in different situations.
As a matter of fact, Appendix B shows that numerous parameters affect the temperature profiles,
including the elastic constant $k^2$ and any perturbation of the interaction forces. In particular,
the presence of different degrees of stochasticity produces different behaviours, and only in the
equilibrium limit may the equivalence of the different models be established.

Independently of the nature of the thermostats, ``realistic'' particle interactions, or mechanisms
preventing the conservation of the most obvious dynamical quantities do not lead to the onset of
local thermodynamic equilibrium \cite{JR10,JR06,DharAP}.
One reason is that certain correlations (e.g.\
particles order) persist in time, violating the molecular chaos hypothesis of kinetic
theory, in 1-d or quasi-1-d systems. Thus, some kind of anomalous behaviour is to be
generically expected, as a consequence of the low dimensionality of the dynamics, rather
than of the peculiarities of thermostats. Indeed, in 3 dimensions,
the same thermostatting mechanisms do not lead to anomalous behaviour.
Even in cases of apparently normal transport in 1-d, this is not as robust, with respect
to variations of the microscopic parameters, as it is
in macroscopic thermodynamic phenomena \cite{JR10,JR06}, indicating that genuine local
thermodynamic equilibrium does not hold in those cases. Experimental evidence, although
presently scanty,
confirms this picture, cf.\ e.g. \cite{DharAP,NatureNano,Chang,JBR08,Siwy,HaMa,Levitt}.
Further study is desired, to understand the relation between the peculiarities of
the 1-d models so far considered in the literature and those of real systems, not in
local thermodynamic equilibrium.

We found the linear relation (\ref{LinRel}) between the particles mean displacements
from their equilibrium positions and the temperature profiles.
Apart from the temporal asymmetries of the fluctuations of the main observables,
considered in Section 3, relation (\ref{LinRel}) is the only result which does
not show a delicate dependence on the details of the microscopic dynamics.
It is robust against all modifications of the dynamics, which we have
investigated, and must therefore play an important role in the identification
of universality classes and in the equivalence of thermostats in 1-d systems

Our analysis of systems with purely harmonic and purely hard core interactions indicates
that chaos, or a generic source of randomness, {\em per se}, does not suffice to establish
standard nonequilibrium steady states. On the other hand, it has been observed by various
authors that dynamical chaos is not even necessary for that, see e.g. \cite{JR06,Squares}.

\newpage

\section*{Appendix A}
To assess the relevance of our results for the large $N$ limit, and to compare with the results
of Refs.\cite{LMMP09,LMMP10} we have produced temperature profiles for hard core collisions
and $k^2=1$, with $T_\ell=320$, $T_r=20$, $\theta_\ell=\theta_r=1$, and growing $N$.
Figure 11 shows these profiles, plotted as functions of the rescaled variable
$x=i/N$. One observes a scaling similar to that of Fig.1 in Ref.\cite{LLP97},
where the asymptotic profile is well approximated, on the scale of the figure, at
moderately large values of $N$.

\vskip 15pt

\centerline{\includegraphics[width=6.5cm,height=6.3cm]{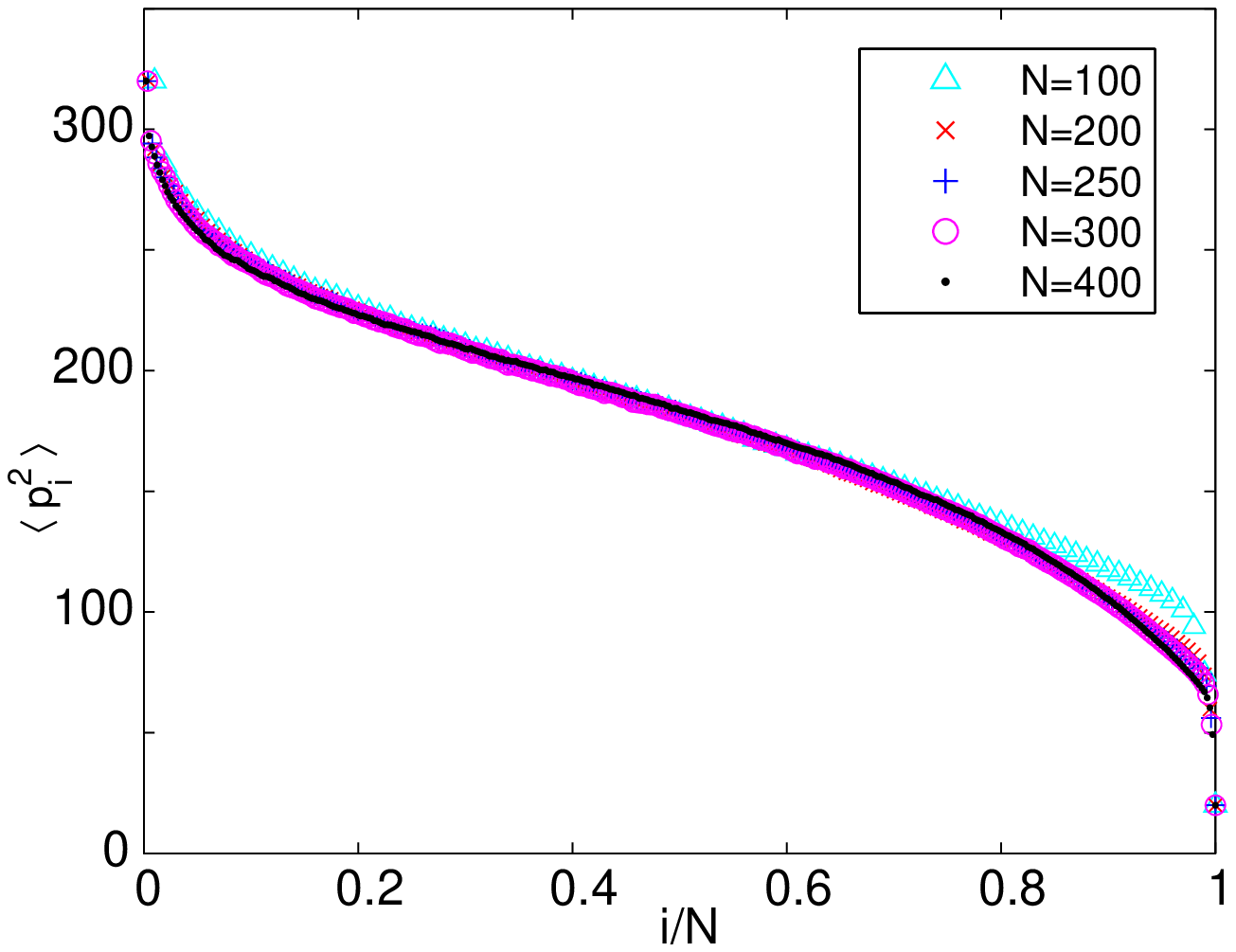}}
\noindent
{\small {\bf Figure 11.} (Color
 online) Temperature profiles as functions of $i/N$, with
$k^2=1$, with $T_\ell=320$, $T_r=20$  $\theta_\ell=\theta_r=1$, for $N=100,200,250,300,400$.
A rather rapid convergence to the asymptotic profile, $T_\infty(y)$, is observed.}

\vskip 15pt

\noindent
Magnifying this figure, one observes that the profile slowest convergence occurs close to the
cold boundary, cf.\ Fig.\ 12.

\vskip 15pt

\centerline{\includegraphics[width=6.5cm,height=6.3cm]{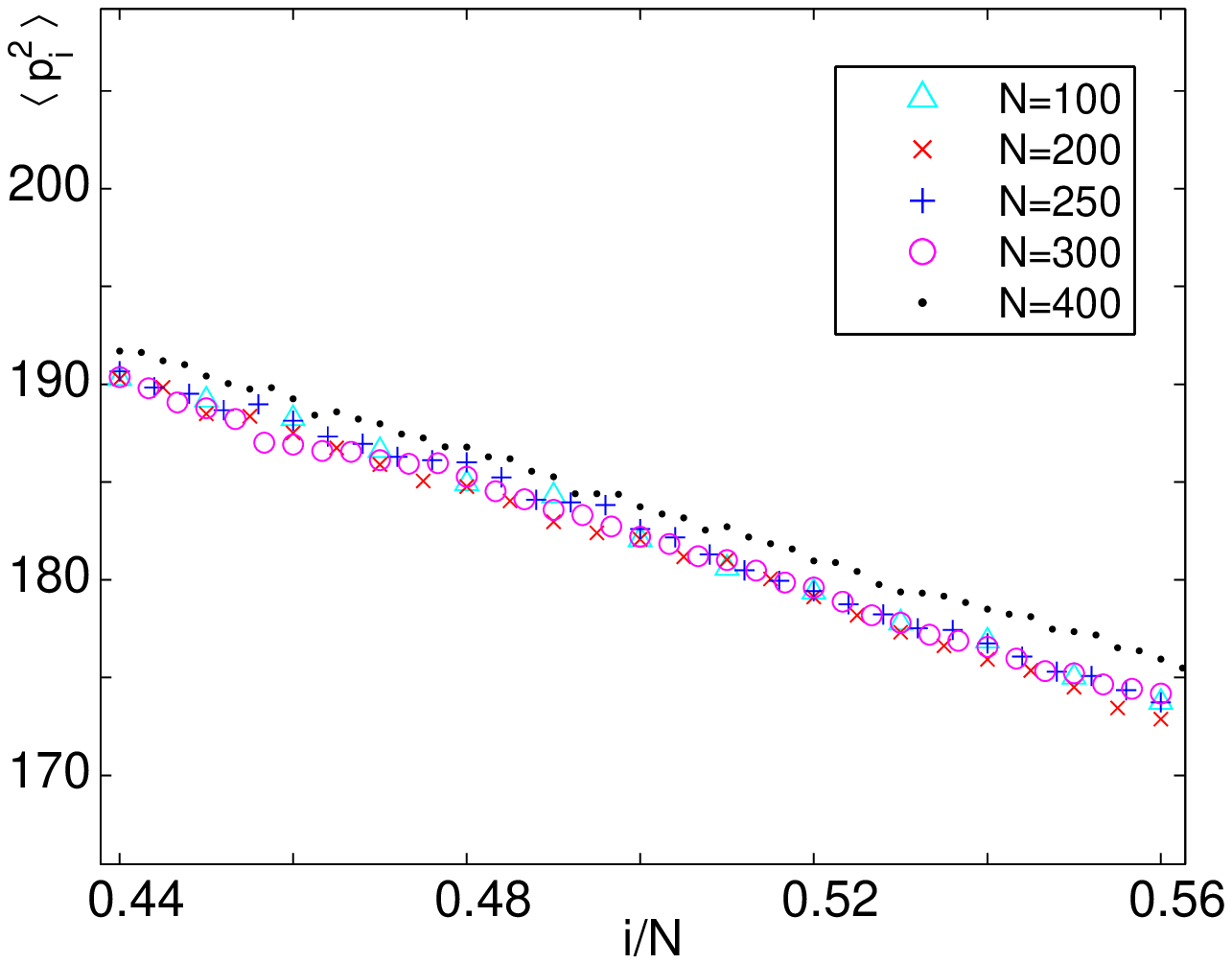} ~~
            \includegraphics[width=6.5cm,height=6.3cm]{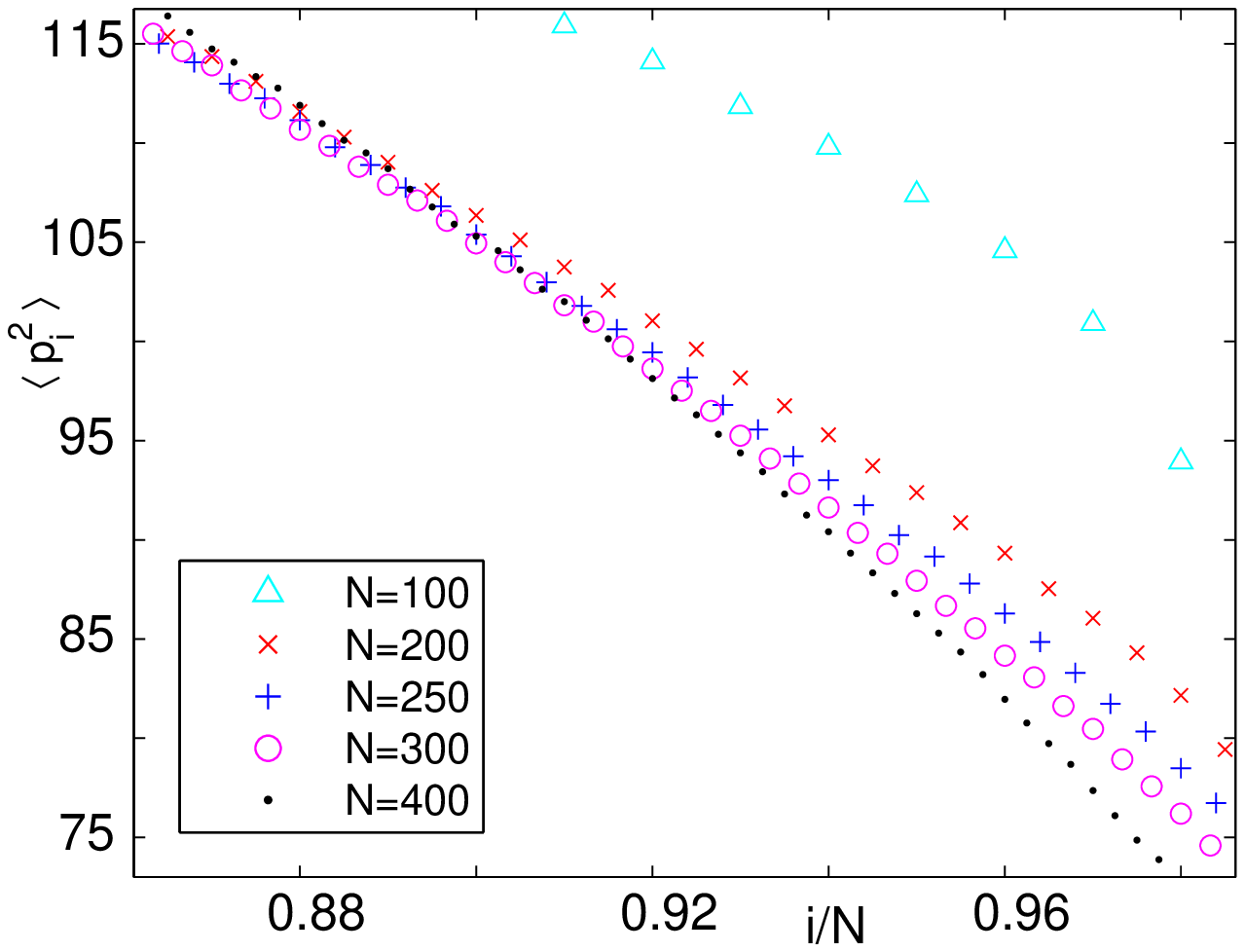}
}
\noindent
{\small {\bf Figure 12.} (Color
 online) Magnification of Fig.\ 11 concerning the center of the chain
(left panel) and the cold side (right panel).}

\vskip 15pt

\noi
The difference $\Delta_N(y)=T(y)-T_N(y)$ between the analytic profile $T$ of Ref.\cite{LMMP09}
and our profiles, $T_N$, with normalized variable $y \in [-1,1]$, for finite $N$, does not
scale with $N$ as the difference between asymptotic and finite $N$ profiles of \cite{LMMP09}.
If the scaling was confirmed, we would have had
\be
\Delta_N(y)\propto N^{-1/3} g(y)
\ee
where $g(y)$ depends on the details of the model, but not on $N$. As a matter of fact, the absolute
value of $\Delta_N(y)$ initially grows with $N$, rather than decreasing, e.g.\ in the center of chain.
This does not prevent the scaling to set in at still larger values of $N$, although this looks
unlikely, given the observed convergence of our profiles to a different asymptotic shape.

\newpage

\section*{Appendix B}
We consider a stochastic perturbation of the model of Section 2. If the dynamics of Section 2
leads two particles to collide at a given instant of time $t$, they will actually collide
with probability $1-p$, $p \in [0,1]$, at time $t$. Taking $p=0$ yields the dynamics of
Section 2, while $p=1$ yields the purely harmonic dynamics of Section 4, because no collision
takes place. For a system of
$N=100$ particles, we find that the temperature profiles strongly depend on the values of the
pair of parameters $p$ and $k^2$, confirming that nonequilibrium chains of oscillators
may hardly be part of a unique universality class, except in the equilibrium limit,
i.e.\ for temperatures at the boundaries very close to each other. The states of these
chains are indeed too far from local equilibrium, and sensitive to many details of
the microscopic dynamics, in general.

In particular, Fig.\ 13 shows the temperature profiles for different values of $p$ and
$k^2 = 0.1$, for two temperature differences $(T_\ell - T_r)$. The dependence
on $p$ is quite strong although, in both cases, the profiles settle around the unperturbed
ones, when $p$ decreases. The cases with lower temperature difference more rapidly collapse
on a unique shape, different from the theoretical $T$ of \cite{LMMP09}.

\vskip 15pt

\centerline{\includegraphics[width=8cm]{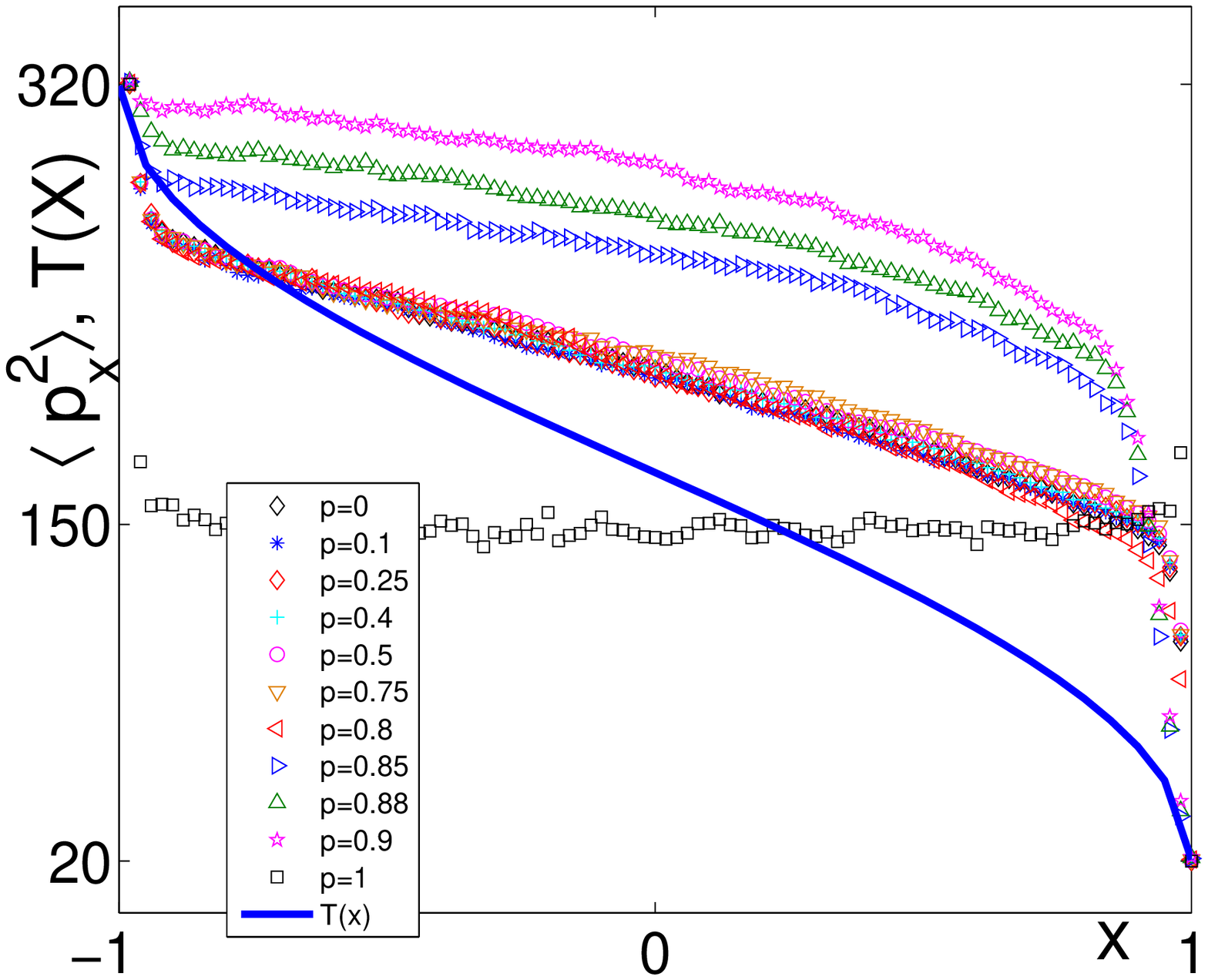} ~~
            \includegraphics[width=8cm]{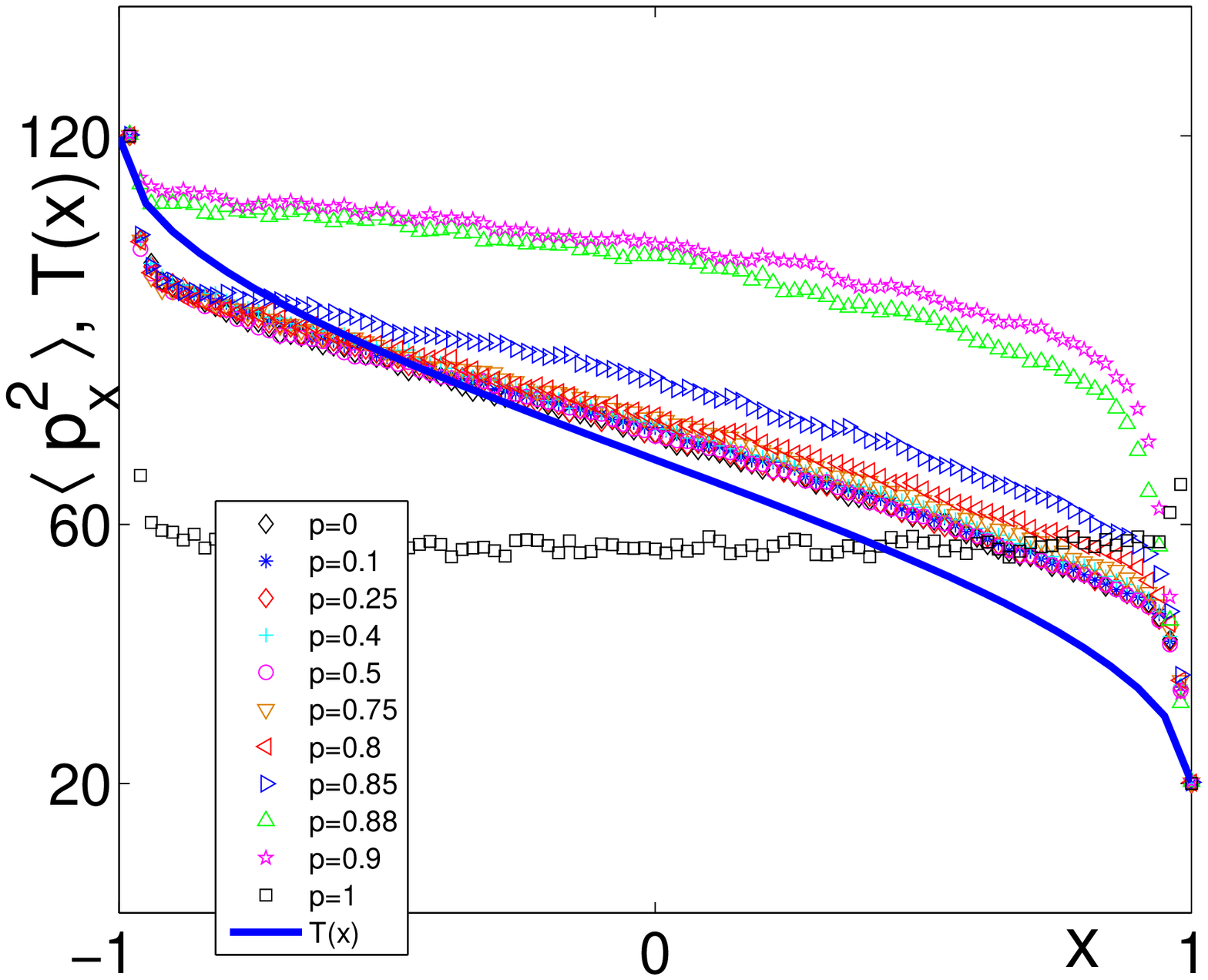}
}
\noindent
{\small {\bf Figure 13.} (Color
 online) Temperature profiles as functions of $p$, for $N=100$, $k^2 = 0.1$.
The left panel reports the case with $T_\ell=320$ and $T_r=20$. The right panel reports
the case with $T_\ell=120$ and $T_r=20$. The continuous line represents the temperature profile $T(x)$ of eq. (\ref{TOFX}).
The lower temperature jump leads to more rapid
convergence to the unperturbed profiles, as $p \to 0$.}

\vskip 15pt

\noi
Figure 14, shows that increasing the rigidity of the chains, reduces the effect of the
parameter $p$. Comparing Figs.\ 13 and 14 leads to the conclusion that,
even for a fixed kind of thermostats, the equivalence of the different microscopic
dynamics requires various parameters to be adjusted, in general. In our case, $k^2$
plays an important role, which it did not in Refs.\cite{LMMP09,LMMP10}. Taking
of $k^2=2$ does not yield profiles which approximate the theoretical
one of \cite{LMMP09} more closely than $k^2=1$.

\vskip 15pt

\centerline{\includegraphics[width=8cm]{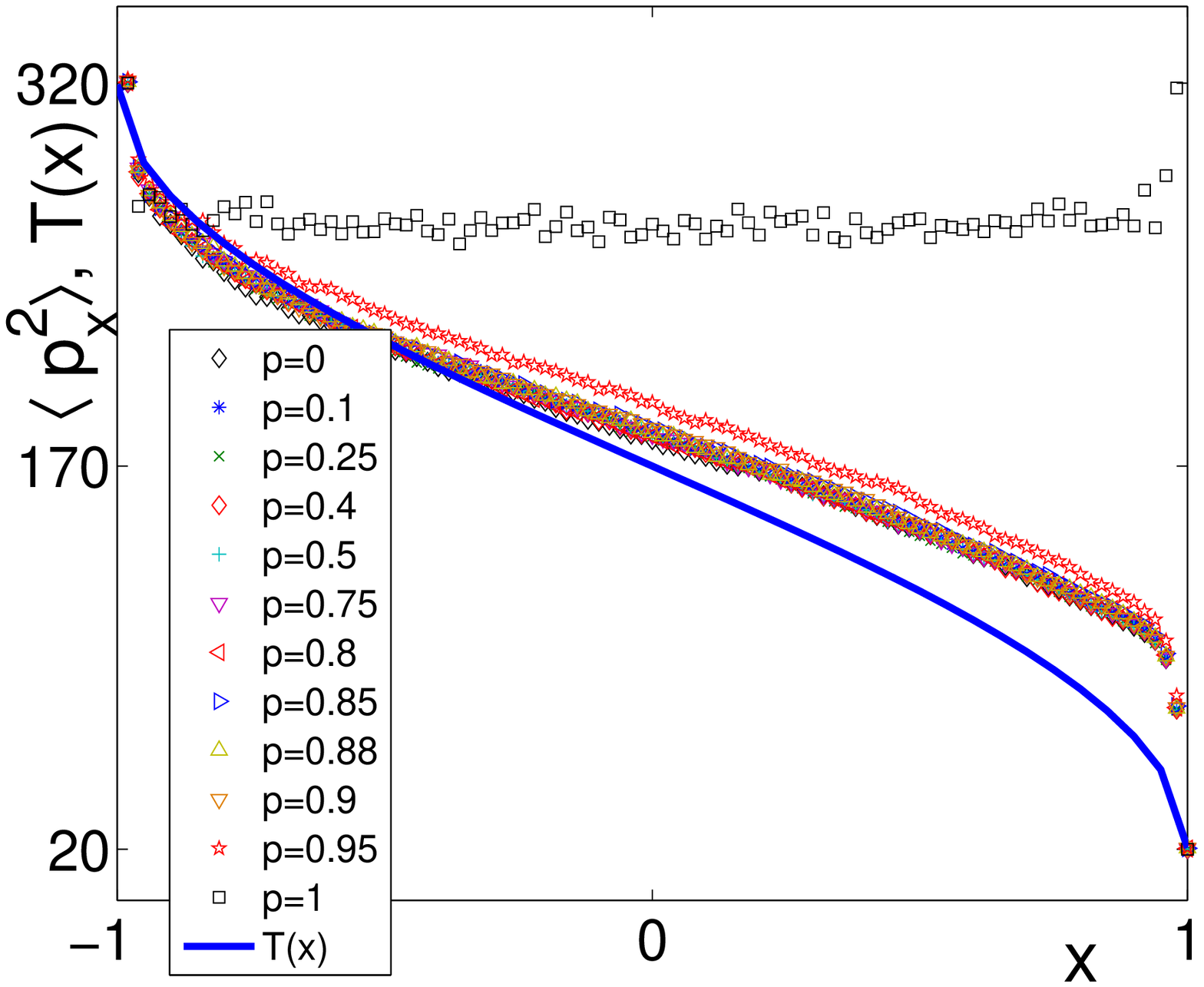} ~~
            \includegraphics[width=8cm]{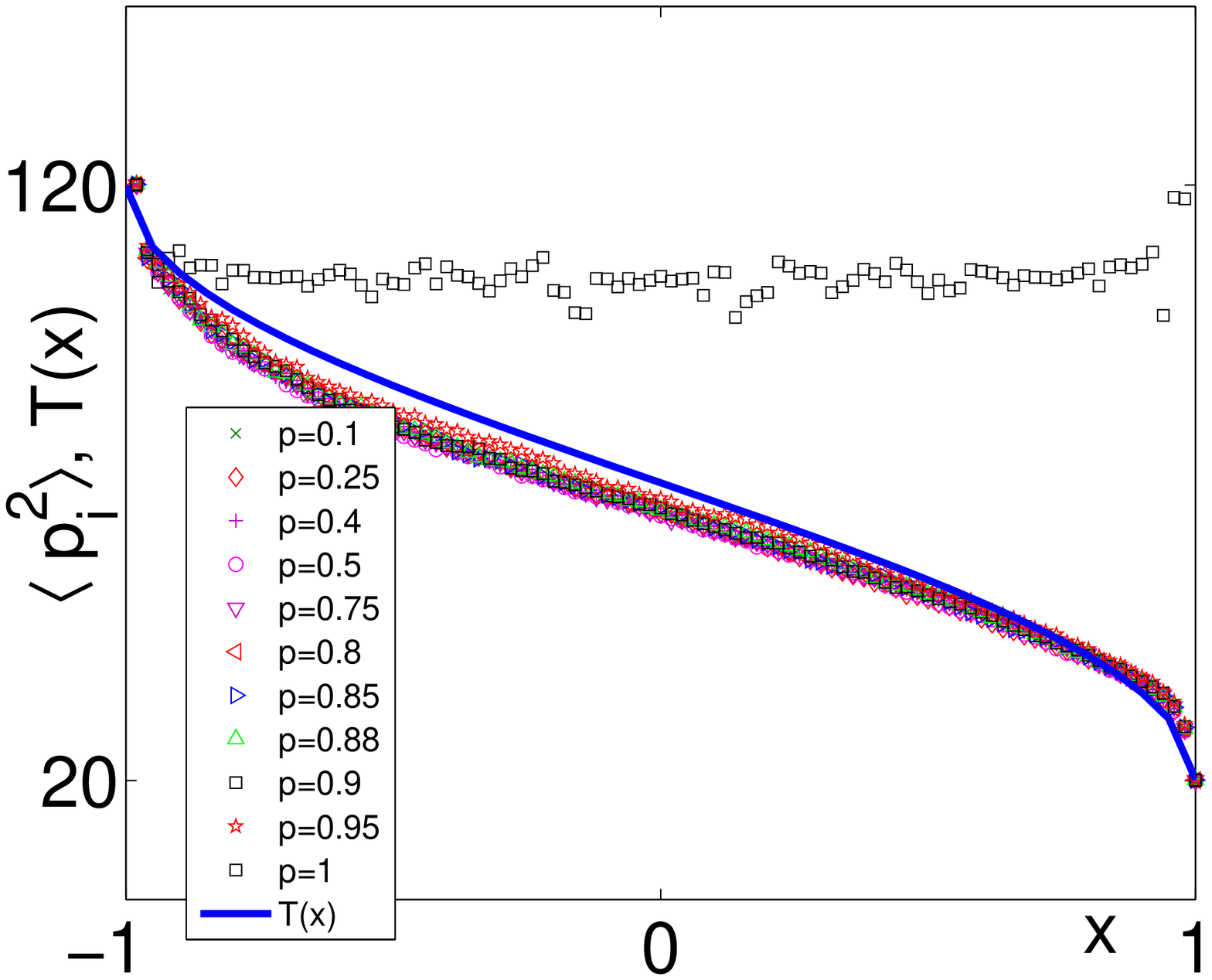}
}
\noindent
{\small {\bf Figure 14.} (Color
 online) Temperature profiles as functions of $p$, for $N=100$, $k^2 = 1$.
The left panel reports the case with $T_\ell=320$ and $T_r=20$. The right panel reports
the case with $T_\ell=120$ and $T_r=20$. The continuous line represents the temperature profile $T(x)$ of eq. (\ref{TOFX}). The dependence on $p$ is reduced at this higher value of $k^2$.}

\vskip 15pt

\noi
The only result which is robust against all the modifications of the dynamics, which we have
investigated, is the validity of the linear relation (\ref{LinRel}), which must then play
an important role in the identification of universality classes and in the equivalence of
thermostats in 1-d systems. Figures 15 and 16 illustrate this fact, confirmed by all
simulations we performed, for only two of these cases.

\vskip 15pt

\centerline{\includegraphics[width=8cm]{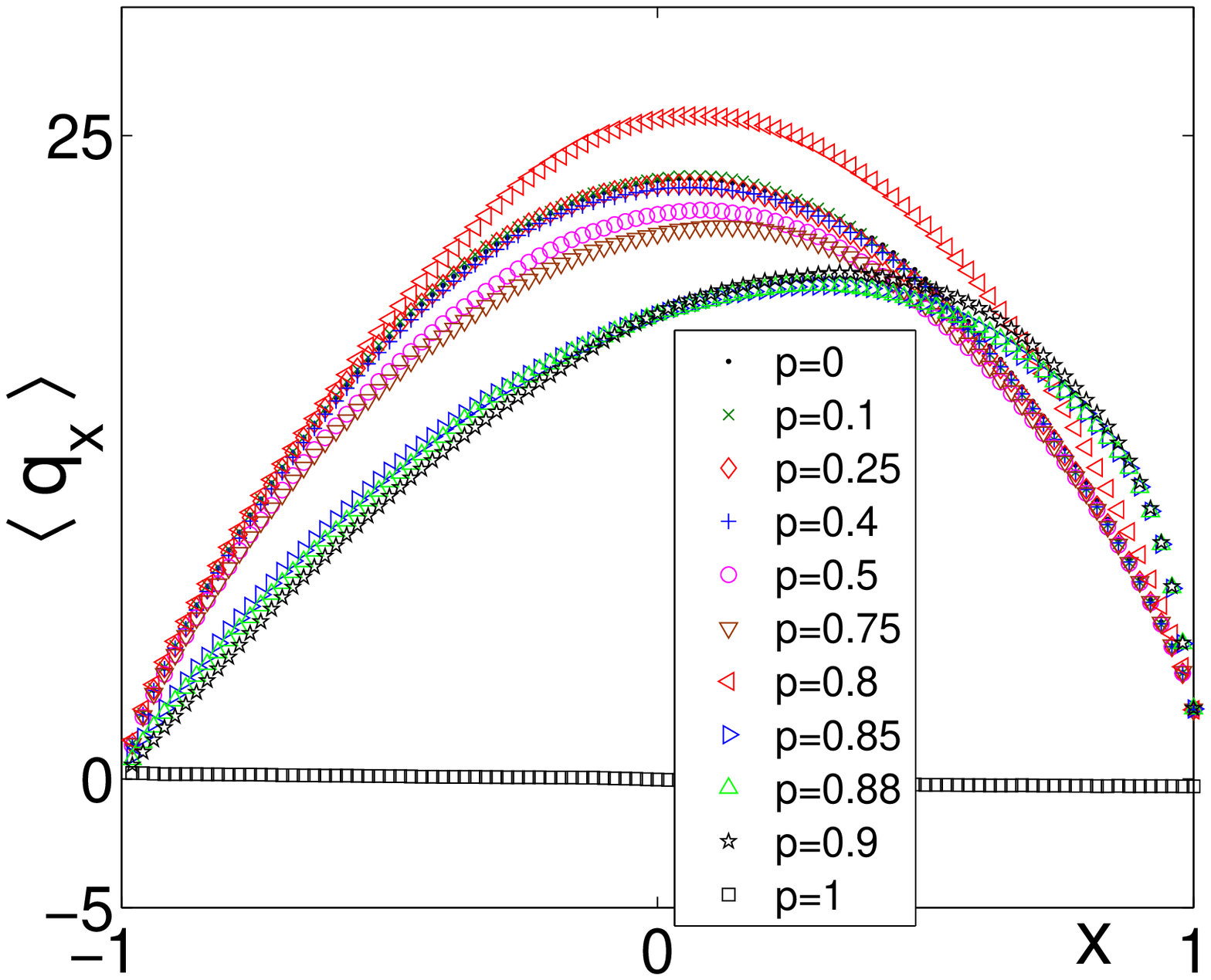} ~~
            \includegraphics[width=8cm]{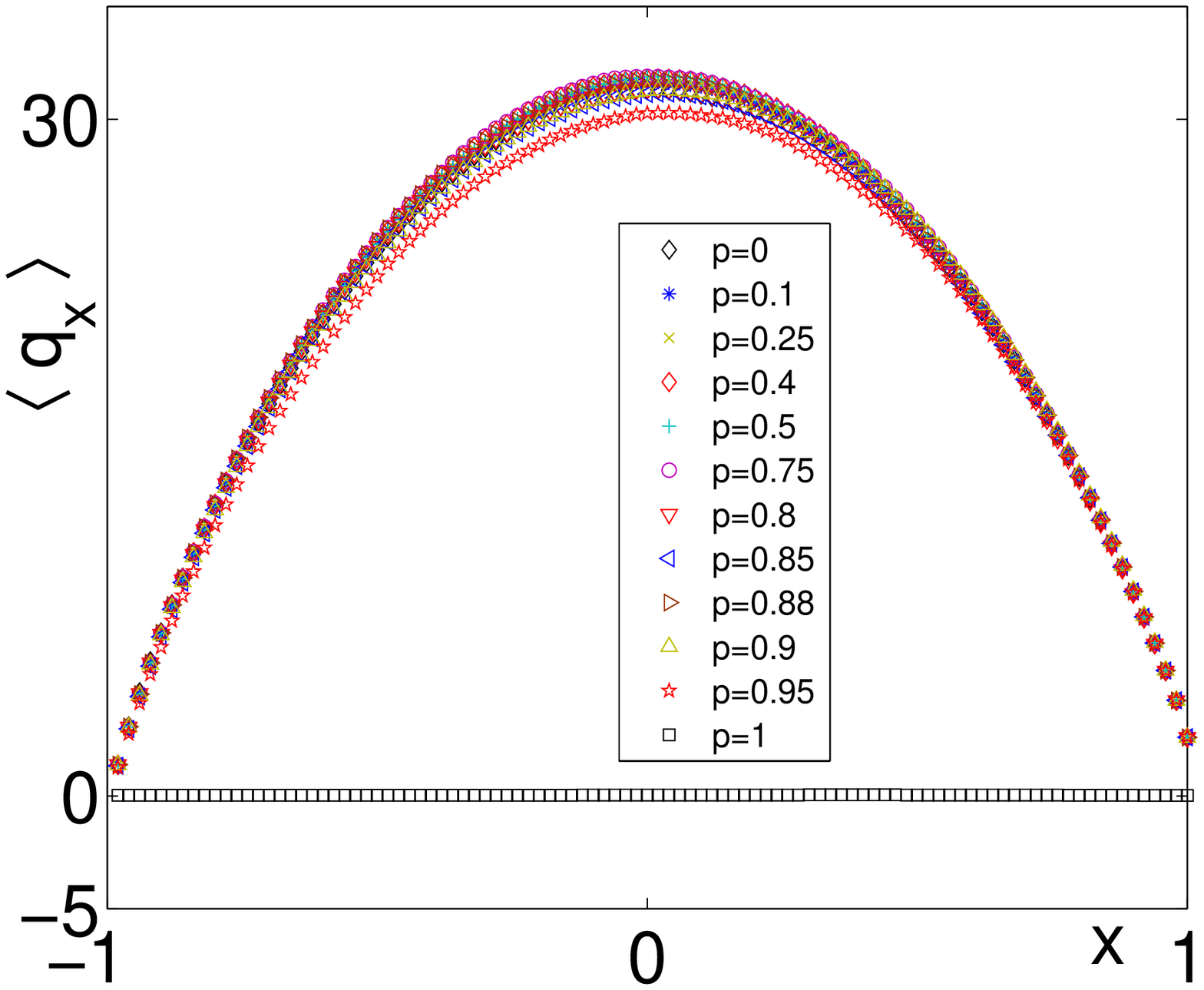}
}
\noi
{\small {\bf Figure 15.} (Color
 online)  Mean displacement $\langle q_i \rangle_{k^2=0.1}$
from the equilibrium centres of oscillation of the particles, $a i$ (left panel) and
mean displacement $\langle q_i \rangle_{k^2=1}$ (right panel), for $N=100$, $T_\ell=320$ and $T_r=20$.
Larger $k^2$ reduces the $p$ dependence, except for $p=0$, which does not
sustain temperature gradients.}

\vskip 15pt

\noi
For a given choice of the parameters of the deterministic model, the values of the
parameters $\zb_1$ and $\zb_2$ of Eq.(\ref{LinRel}) do not depend on $p$, as long as
$p < 0.9$. They change discontinuosly for $p \ge 0.9$, i.e.\ close to the purely
harmonic chains, which are quite peculiar models, compared to the others.

\vskip 15pt

\centerline{\includegraphics[width=8cm]{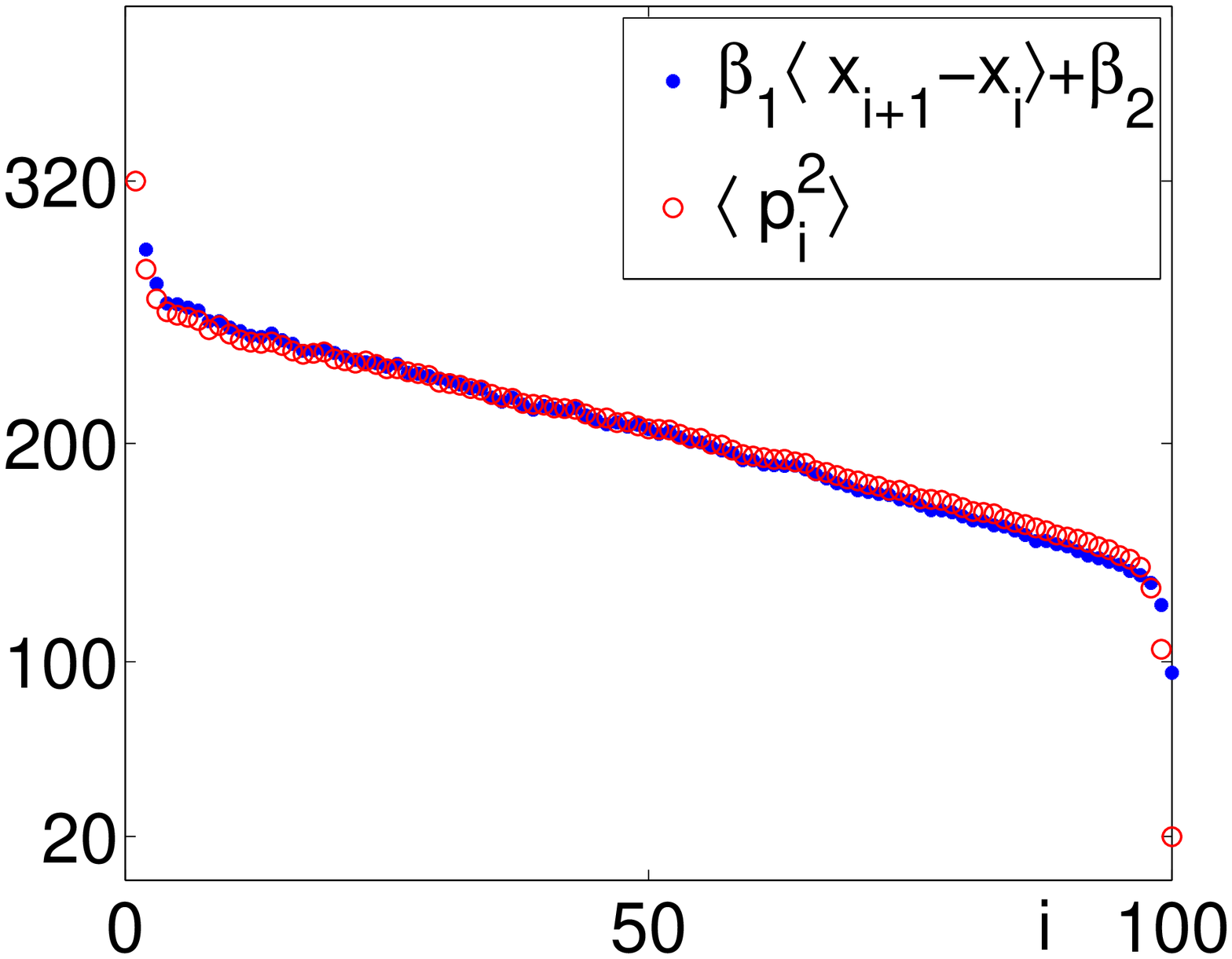} ~~
            \includegraphics[width=8cm]{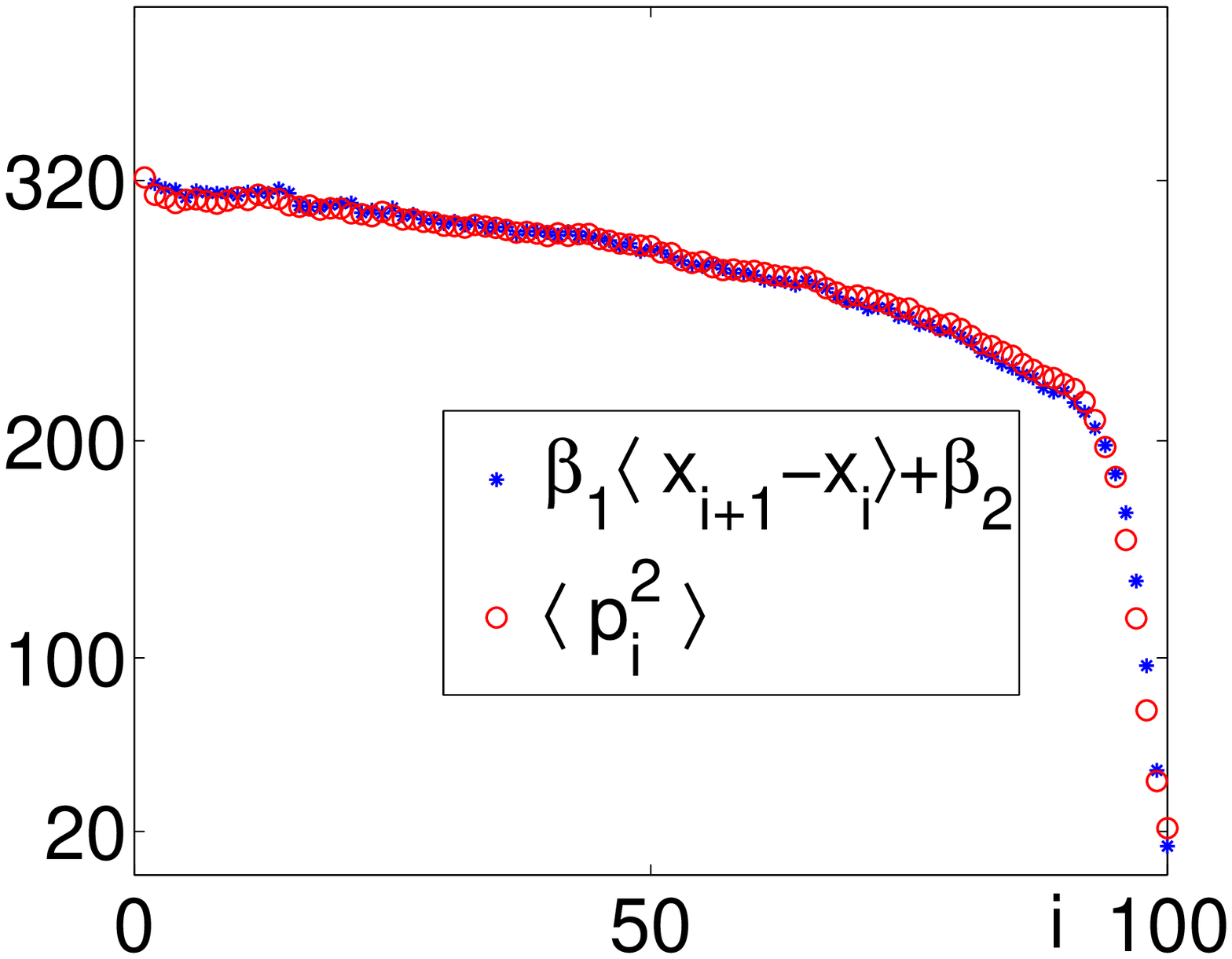}
}
\noi
{\small {\bf Figure 16.} (Color
 online)
Relation between mean coordinates difference
$\langle x_{i+1} - x_i \rangle$ and temperature profiles, for $N=100$, $T_\ell=320$,
$T_r=20$, $k^2=0.1$. The left panel
reports the case $p=0.1$, the right panel reports the case $p=0.9$.
Relation (\ref{LinRel}) is confirmed in all instances as discussed in Section 2.
}

\newpage

\section*{Acknowledgements}
The authors are grateful to S.\ Lepri, C.\ Mej\'{i}a Monasterio and A.\ Politi for
enlightening discussions and for reading a preliminary draft of this paper. LR
gratefully acknowledges financial support
from the European Research Council under the European Community's Seventh Framework
Programme (FP7/2007-2013) / ERC grant agreement n 202680. The EC is not liable for
any use that can be made on the information contained herein.

%%%%%%%%%%%%%%%%%%%%%%%%%%%%

\end{document}